\documentclass[a4paper,11pt]{article}
\usepackage{jcappub}
\usepackage{amsmath}
\usepackage{graphicx}

\makeatletter
\gdef\@fpheader{}
\makeatother

\newlength{\fullw}
\newlength{\halfw}
\newlength{\threefigw}
\newlength{\twofigw}
\newlength{\onefigw}
\newcommand{\order}[1]{\mathcal{O}\!\left(#1\right)}
\newcommand{\boldmathsymbol}[1]{{\ensuremath{\boldsymbol{#1}}}}
\newcommand{\sss}[1]{{\scriptscriptstyle{#1}}}

\newcommand{\heaviside}[1]{\mathrm{\Theta}\!\left( #1 \right)}
\newcommand{\heavisideb}[1]{\mathrm{\Theta}\!\left[ #1 \right]}

\newcommand{\vect}[1]{\boldmathsymbol{#1}}
\newcommand{\negleft}{\negthinspace\left}
\newcommand{\dirac}[1]{\delta\negleft(#1\right)}
\newcommand{\diracb}[1]{\delta\negleft[#1\right]}
\newcommand{\ddprime}[1]{\mathaccent"707D{#1}}

\newcommand{\MULTINEST}{\texttt{MULTINEST}}

\newcommand{\Ho}{H_0}
\newcommand{\tnow}{t_0}
\newcommand{\chio}{\chi_0}

\newcommand{\Hz}{\mathrm{Hz}}
\newcommand{\nHz}{\mathrm{nHz}}

\newcommand{\ud}{\mathrm{d}}
\newcommand{\uc}{\mathrm{c}}

\newcommand{\uk}{\mathrm{k}}
\newcommand{\ukk}{\mathrm{kk}}
\newcommand{\us}{\mathrm{s}}

\newcommand{\ueq}{\mathrm{eq}}
\newcommand{\ugw}{\mathrm{gw}}

\newcommand{\usgw}{\mathrm{sgw}}

\newcommand{\urad}{\mathrm{rad}}

\newcommand{\uobs}{\mathrm{obs}}
\newcommand{\ucrit}{\mathrm{crit}}
\newcommand{\ubeam}{\mathrm{beam}}
\newcommand{\uknee}{\mathrm{knee}}
\newcommand{\upeak}{\mathrm{peak}}
\newcommand{\ucut}{\mathrm{cut}}
\newcommand{\uem}{\mathrm{em}}

\newcommand{\uM}{\mathrm{M}}
\newcommand{\uR}{\mathrm{R}}

\newcommand{\calF}{\mathcal{F}}
\newcommand{\calC}{\mathcal{C}}
\newcommand{\calR}{\mathcal{R}}
\newcommand{\tfourFo}{\tnow^4\calF_0}

\newcommand{\Xd}{{\dot{X}}}

\newcommand{\Xp}{\acute{X}}
\newcommand{\Xpp}{\ddprime{X}}

\newcommand{\bX}{\vect{X}}
\newcommand{\bx}{\vect{x}}
\newcommand{\br}{\vect{r}}
\newcommand{\bkappa}{\vect{\kappa}}

\newcommand{\bnhat}{\vect{\hat{n}}}

\newcommand{\bXd}{\dot{\bX}}
\newcommand{\bXp}{\acute{\bX}}

\newcommand{\bXpp}{\ddprime{\bX}}

\newcommand{\etas}{\eta_{\sss{\us}}}
\newcommand{\ssaddle}{\sigma^{\sss{(\us)}}}
\newcommand{\tsaddle}{\tau^{\sss{(\us)}}}
\newcommand{\skink}{\sigma^{\sss{(\uk)}}}
\newcommand{\tkink}{\tau^{\sss{(\uk)}}}

\newcommand{\Nkink}{N_{\sss{\uk}}}
\newcommand{\Ncusp}{N_{\sss{\uc}}}
\newcommand{\Ncoll}{N_{\sss{\uk\uk}}}

\newcommand{\rhocrit}{\rho_{\ucrit}}

\newcommand{\chimat}{\chi_{_\uM}}
\newcommand{\chirad}{\chi_{_\uR}}

\newcommand{\OmegaR}{\Omega_{\urad}}

\newcommand{\OmegaL}{\Omega_{\Lambda}}

\newcommand{\OmegaGW}{\Omega_{\ugw}}
\newcommand{\OmegaSGW}{\Omega_{\usgw}}
\newcommand{\OmegaSGWH}{\hat{\Omega}_{\usgw}}

\newcommand{\gammad}{\gamma_\ud}
\newcommand{\gammac}{\gamma_\uc}

\newcommand{\gammai}{\gamma_\infty}
\newcommand{\gammaw}{\gamma_{1}}
\newcommand{\gammas}{\gamma_{\star}}

\newcommand{\ellc}{\ell_\uc}
\newcommand{\elld}{\ell_\ud}

\newcommand{\zeq}{z_\ueq}

\newcommand{\zx}{z_{\times}}
\newcommand{\zs}{z_{\star}}
\newcommand{\zw}{z_{1}}
\newcommand{\zwc}{z_{\uc}}
\newcommand{\zcc}{z_{\Lambda}}

\newcommand{\rdof}{\mathcal{Q}}
\newcommand{\gzero}{g_\sss{0}}
\newcommand{\g}{g}
\newcommand{\gs}{q}
\newcommand{\gszero}{\gs_\sss{0}}

\newcommand{\omegaknee}{\omega_{\sss{\uknee}}}
\newcommand{\omegapeak}{\omega_{\sss{\upeak}}}
\newcommand{\omegac}{\omega_{\uc}}

\newcommand{\GU}{GU}
\newcommand{\GUU}{GU^2}
\newcommand{\Pgw}{P_\ugw}

\newcommand{\ncut}{n_{\sss{\ucut}}}

\newcommand{\hit}{\bar{h}}
\newcommand{\hitlwz}{\hit_{\flat}}
\newcommand{\hitrms}{\hit_{\uc}}
\newcommand{\hitstar}{\hit_{\star}}
\newcommand{\hitt}{\tilde{h}}
\newcommand{\hittstar}{\hitt_{\star}}
\newcommand{\hith}{\hat{h}}
\newcommand{\hithstar}{\hith_{\star}}

\newcommand{\thetab}{\theta_{\ubeam}}
\newcommand{\Deltab}{\Delta_{\ubeam}}
\newcommand{\Omegab}{\Omega_{\ubeam}}

\newcommand{\zem}{z_\uem}

\newcommand{\aem}{a_\uem}
\newcommand{\chiobs}{\chi_\uobs}
\newcommand{\chiem}{\chi_\uem}
\newcommand{\etaobs}{\eta_\uobs}

\newcommand{\rhogw}{\rho_\ugw}

\newcommand{\Cbar}{\bar{\calC}}
\newcommand{\Cbarstar}{\Cbar_{\star}}
\newcommand{\Ctilde}{\tilde{\calC}}
\newcommand{\Ctildestar}{\Ctilde_{\star}}
\newcommand{\Chat}{\hat{\calC}}

\newcommand{\fref}{f_0}

\title{Stochastic gravitational waves from cosmic string loops in scaling}

\author[a]{Christophe Ringeval}
\author[b]{and Teruaki Suyama}

\affiliation[a]{Centre for Cosmology, Particle Physics and Phenomenology,
  Institute of Mathematics and Physics, Louvain University, 2 Chemin
  du Cyclotron, 1348 Louvain-la-Neuve, Belgium}

\affiliation[b]{Research Center for the Early Universe (RESCEU),
  University of Tokyo, Hongo, 7-3-1 Bunkyo-ku, Tokyo 113-0033, Japan}

\emailAdd{christophe.ringeval@uclouvain.be}
\emailAdd{suyama@resceu.s.u-tokyo.ac.jp}

\date{today}

\begin{document}

\abstract{If cosmic strings are formed in the early universe, their
  associated loops emit gravitational waves during the whole cosmic
  history and contribute to the stochastic gravitational wave
  background at all frequencies. We provide a new estimate of the
  stochastic gravitational wave spectrum by considering a realistic
  cosmological loop distribution, in scaling, as it can be inferred
  from Nambu-Goto numerical simulations. Our result takes into account
  various effects neglected so far. We include both gravitational wave
  emission and backreaction effects on the loop distribution and show
  that they produce two distinct features in the spectrum. Concerning
  the string microstructure, in addition to the presence of cusps and
  kinks, we show that gravitational wave bursts created by the
  collision of kinks could dominate the signal for wiggly strings, a
  situation which may be favoured in the light of recent numerical
  simulations. In view of these new results, we propose four
  prototypical scenarios, within the margin of the remaining
  theoretical uncertainties, for which we derive the corresponding
  signal and estimate the constraints on the string tension put by
  both the LIGO and European Pulsar Timing Array (EPTA)
  observations. The less constrained of these scenarios is shown to
  have a string tension $\GU \le 7.2 \times 10^{-11}$, at $95\%$ of
  confidence. Smooth loops carrying two cusps per oscillation verify
  the two-sigma bound $\GU \le 1.0 \times 10^{-11}$ while the most
  constrained of all scenarios describes very kinky loops and
  satisfies $\GU \le 6.7\times 10^{-14}$ at $95\%$ of confidence.}

\keywords{Cosmic Strings, Gravitational Waves, Loops}

\maketitle

\section{Introduction}
\label{sec:intro}

Cosmic strings are topological line defects expected to be formed
during phase transitions in the early Universe~\cite{Kirzhnits:1972,
  Kibble:1976}. They could also be mimicked by more fundamental
objects, such as D1-brane of F-strings, directly stemming from String
Theory and referred to as cosmic superstrings~\cite{Witten:1985fp,
  Dvali:1998pa}. In the cosmological context, once formed, a string
network relaxes towards a cosmological attractor that exhibits
universal statistical properties~\cite{Hindmarsh:1994re, Durrer:2002,
  Polchinski:2004ia, Davis:2008dj, Copeland:2009ga,
  Sakellariadou:2009ev, Ringeval:2010ca}. For instance, the number of
strings crossing a Hubble radius remains stationary for the rest of
the cosmic history, this is the so-called ``scaling'' regime. For
Nambu-Goto strings, this implies that all observable predictions
depend on one parameter only which is the string energy density $U$
(also equals to the string tension), a quantity related to the energy
scale at which the strings have been formed.
The determination of a Nambu-Goto network's scaling properties in
Friedmann-Lema\^itre-Robertson-Walker spacetime (FLRW) is a
non-trivial issue and has been the subject of numerical investigations
in the last thirty years~\cite{Albrecht:1989, Bennett:1989,
  Allen:1990, Bennett:1990}. These works have shown that the long
strings are indeed able to reach the cosmological attractor by loosing
energy under the form of loops, these ones loosing energy in turn
under the form of smaller loops and gravitational waves (GW). As a
result, it has been known for a long time that cosmic strings could
potentially be a dominant source for the Universe stochastic
gravitational wave background~\cite{Vilenkin:1981, Hogan:1984is,
  Accetta:1988bg, Sakellariadou:1990ne, Bennett:1990ry}. This picture
could however be challenged by the Abelian Higgs strings in which
strings can decay by particle emission instead of
GW~\cite{Vincent:1998, Moore:2002, Hindmarsh:2008dw,
  Hindmarsh:2017qff}. In the following, we will be focused on
Nambu-Goto strings only although our modelling could be extended to
include decay channels others than GW.

In an expanding Universe, the distribution of cosmic string loops,
i.e. the number density of loops with respect to their size, reaches a
scaling regime, a result first shown in Ref.~\cite{Ringeval:2005kr}
and soon after confirmed in Refs.~\cite{Vanchurin:2005yb,
  Martins:2005es}. However, most of the works having estimated the
stochastic gravitational wave spectrum from loops have assumed utterly
simplified loop distribution, or loop production function, usually
postulated to be a Dirac function peaked at a length equal to some
given fraction of the horizon size~\cite{Caldwell:1991jj,
  Damour:2000wa, DePies:2007bm, Regimbau:2011bm, Binetruy:2012ze,
  Kuroyanagi:2012wm, Aasi:2013vna, Henrot-Versille:2014jua,
  Sousa:2016ggw}. In the meanwhile, more recent Nambu-Goto simulations
presented in Ref.~\cite{Blanco-Pillado:2013qja} have independently
recovered the power-law shape of the loop distribution originally
found in Ref.~\cite{Ringeval:2005kr} thereby giving a robust picture
of the loop scaling regime on the length scales reachable in numerical
simulations. The authors of Ref.~\cite{Blanco-Pillado:2013qja} have
also provided a new estimation of the gravitational wave spectrum
stemming from their loop distribution. As we discuss in
section~\ref{sec:pw}, when the additional effects we are considering
are switched off, our spectrum is of similar shape and amplitude
compared to the one of Ref.~\cite{Blanco-Pillado:2013qja}.

The shape of the loop distribution on the length scales probed by
numerical simulations is however not enough to uniquely determine
their associated stochastic GW spectrum. Some theoretical
uncertainties remain.

Firstly, gravitational wave \emph{emission} impacts the scaling loop
distribution. All loops shrink by loosing energy under the form of GW
and the loop distribution ends up being modified when such a process
becomes faster than the other mechanisms at work in a string network
(production from string self-intersections and loop
fragmentation). The emitted GW power is given by $\Pgw = \Gamma \GUU$
such that GW evaporation dominates for loops of size $\ell < \Gamma
\GU t$, $G$ being the Newton constant, $t$ the cosmic time and
$\Gamma$ a numerical constant estimated to be $\Gamma =
\order{50}$~\cite{Vachaspati:1984gt, PhysRevD.45.1898}. For this
reason, this length scale is the one under which numerical simulations
cannot be trusted for cosmological purposes. For the maximal allowed
values of $GU = 10^{-7}$, this regime appears for loops smaller than a
millionth the size of the horizon~\cite{Ade:2013xla, Lazanu:2014eya,
  Lizarraga:2016onn}. In the following, we define
\begin{equation}
  \gammad \equiv \Gamma \GU,
\label{eq:gammad}
\end{equation}
the gravitational wave \emph{emission} length scale, measured in unit
of $t$.

Secondly, the gravitational wave signal emitted by a single loop
mostly depends on the string microstructure, and, as shown by Damour
and Vilenkin in Ref.~\cite{Damour:2001bk}, the spectrum at high
frequency is dominated by the transient appearance of ``cusps'' in
the shape of the string, and more generally on piece of strings
approaching the speed of light~\cite{Stott:2016loe}. Its amplitude scales as
$\omega^{-4/3}$, $\omega$ being the GW angular frequency. As a result, the
signal amplitude depends on the number of cusps appearing per loop
oscillation. GW can also be produced from ``kinks'' in the shape of
the string, with an amplitude varying however as
$\omega^{-5/3}$~\cite{Damour:2001bk}. Therefore, kinks are expected to
be a sub-dominant contribution provided their number remains less, or
comparable, to the number of cusps. Unfortunately, the number of cusps
and kinks per loop oscillation is not yet known and cannot be
straightforwardly derived from numerical simulations. These numbers
indeed depend on the so-called gravitational wave backreaction, an
effect which is not included in the simulations. Kinks appear in pairs
from string collisions, then propagate at the speed of light in
opposite directions, and have a tendency to accumulate over the
cosmological evolution. All Nambu-Goto simulations have shown that the
loops are extremely wiggly, i.e. filled with kinks~\cite{Bennett:1989,
  Allen:1990}. However, by emitting gravitational waves, one expects
backreaction to wash-out the string microstructure under some given
length scale thereby rendering the kinky loops smoother, a mechanism
known to favour the appearance of
cusps~\cite{Blanco-Pillado:2015ana}. This smoothing mechanism is the
standard lore but has been recently challenged in
Refs.~\cite{Wachter:2016hgi, Wachter:2016rwc} in which the effect of
gravitational wave backreaction on the shape of the loops has been
explicitly estimated for various loop shapes. Some loops become
smoother, as expected, but others can actually remain filled with
kinks.

Thirdly, in addition to modify the number of kinks and cusps per loop
oscillations, gravitational wave \emph{backreaction} can also modify
the scaling loop distribution. Because very small loops are created
from the microstructure of larger strings and larger loops,
gravitational wave backreaction is expected to switch off loop
production mechanisms in a string network under a given length
scale. Another simplifying assumption made in all the works having
estimated the stochastic gravitational wave spectrum so far is that
the scale of gravitational wave \emph{backreaction} matches the scale
of gravitational wave \emph{emission}, $\gammad$. However, this is
usually not the case. As shown in Refs.~\cite{Siemens:2001dx,
  Siemens:2002dj, Polchinski:2007rg}, gravitational wave backreaction
on kinky strings may appear at a length scale $\ellc$, which can be
much smaller than $\elld = \gammad t$. In the following, we denote by
$\gammac \equiv \ellc/t$ the backreaction length scale, measured in
unit of $t$. From Ref.~\cite{Polchinski:2007rg}, it has been estimated
to be
\begin{equation}
\gammac \equiv \Upsilon (\GU)^{1+2\chi},
\label{eq:gammac}
\end{equation}
in which $\Upsilon = \order{20}$. The parameter $\chi$ is related to
the tangent vector correlations along the strings and can be read-off
from the shape of the loop distribution on large scales, and thus can
be numerically determined from Nambu-Goto simulations\footnote{The same
parameter $\chi$ enters in the shape of CMB trispectrum generated by
cosmic strings~\cite{Hindmarsh:2009es}.}.

\begin{figure}
  \begin{center}
    \includegraphics[width=\onefigw]{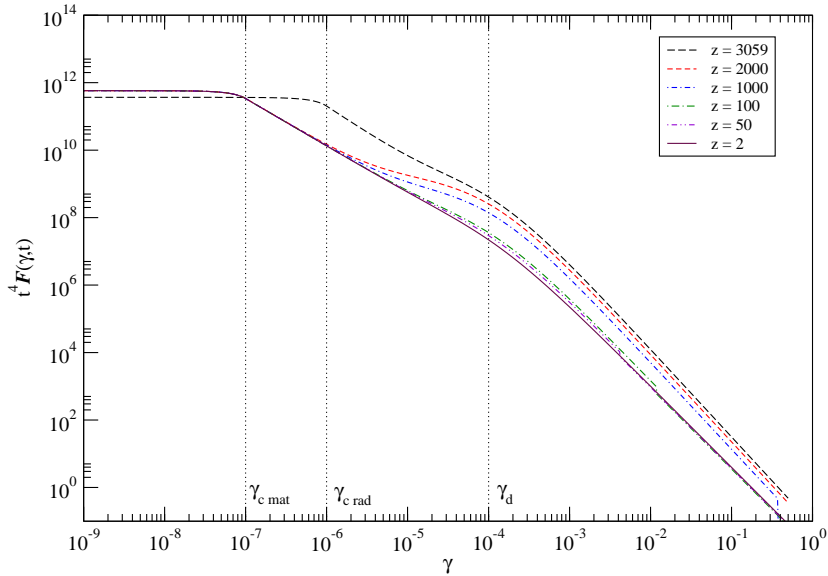}
    \caption{Relaxation of the loop distribution $t^4 \calF(\gamma,z)$
      from its scaling shape in the radiation era (black dashed line
      at redshift $z=3059$) towards the matter era attractor (solid
      line at redshift $z=2$). The values for $\gammac$ in the matter
      and radiation era have been arbitrarily chosen for illustration
      purposes.}
    \label{fig:ldtrans}
  \end{center}  
\end{figure}

In this work, we tackle various of these remaining theoretical
uncertainties and provide a new estimate of the stochastic
gravitational wave spectrum generated by cosmic string loops. We
include both the effect of gravitational wave emission and
gravitational wave backreaction on the scaling loop distribution. In
order to do so, we have followed the method of
Refs.~\cite{Lorenz:2010sm, Peter:2013jj} and we solve a
two-dimensional Boltzmann equation to determine the loop number
density distribution $\calF$ at any redshift $z$:
\begin{equation}
\calF(\gamma,z) \equiv \dfrac{\ud n}{\ud \ell}\,, \qquad \gamma(\ell,z)
\equiv \dfrac{\ell}{t(z)}\,,
\end{equation}
$n$ being the number of loops per unit volume. This Boltzmann equation
includes gravitational wave emission and a scaling loop production
function inferred from the Nambu-Goto simulations of
Ref.~\cite{Ringeval:2005kr}. As shown in Ref.~\cite{Rocha:2007ni},
over the length scales accessible within Nambu-Goto numerical
simulations, this loop production function matches the one predicted
by Polchinski and Rocha in Refs.~\cite{Polchinski:2006ee,
  Dubath:2007mf}. Therefore, we use the Polchinski-Rocha (PR) model to
extrapolate the loop production function down to the scales at which
gravitational wave backreaction appears. As discussed in
Ref.~\cite{Lorenz:2010sm}, the details on how backreaction smooths
the strings is irrelevant for the number of loops, only the scale at
which gravitational backreaction shows up, i.e. the value of $\gammac$,
ends up having an observable effect. This matching with the PR model
allows us to unambiguously determine the parameter $\chi$ of
Eq.~\eqref{eq:gammac}. One gets
\begin{equation}
\chimat = 0.295^{+0.03}_{-0.04}, \qquad \chirad = 0.200^{+0.07}_{-0.10}\,,
\end{equation}
for the matter and radiation era, respectively. The central value is
the best fit to the PR model while the quoted errors are an overly
safe estimate of all the possible numerical systematics (see
Ref.~\cite{Ringeval:2005kr} for the details). In the following, we
will be using only the best fit values for $\chi$ (see also
Ref.~\cite{Hindmarsh:2008dw} for an estimation of $\chi$ for Abelian
Higgs strings). Another advantage of solving a Boltzmann equation for
the loop distribution concerns the transition from the radiation era
to the matter era as we are able to include any relaxation effects on
the loop distribution. The loop distribution at various redshifts
around the transition has been represented in
figure~\ref{fig:ldtrans}. For completeness, we have included the
thermal history effects on the expansion rate of the Universe as these
ones have been shown to affect the spectrum at high
frequencies~\cite{Binetruy:2012ze}. For this purpose, we have used the
effective number of relativistic degrees of freedom derived in
Ref.~\cite{Hindmarsh:2005ix} while considering that the loop scaling
parameters remain unaffected (which is a reasonable assumption).

Concerning the uncertainties associated with the number of kinks and
cusps present on the loops, we propose various well motivated
scenarios that may be viewed as the remaining theoretical errors on
the spectrum. We discuss scenarios having a smooth microstructure and
only two cusps, others having a number of kinks ranging from zero to
$10^2$. For the latter, we show that a new source of gravitational
waves on the string could dominate the spectrum: the collisions of
left-moving and right-moving kinks. Although the collision amplitude decays
faster with frequency than cusps and kinks, it scales as
$\omega^{-2}$, kink collisions emit gravitational waves in all
directions and the number of event per loop oscillations
increases as the square of the number of kinks. In all these
scenarios, the number of cusps, kinks and collisions is bounded from
above as the total power emitted can never exceed $\Pgw = \Gamma
\GUU$. As a result, one can completely explore the range of the
remaining theoretical uncertainties.

The paper is organized as follows. In section~\ref{sec:src} we briefly
introduce our notation before deriving the main equations for
computing the stochastic gravitational wave spectrum. We do not give
excessive details on GW emitted by one cusp and one kink as we follow
in all points the Damour and Vilenkin
calculations~\cite{Damour:2001bk, Siemens:2006vk}. However, some
intermediate results have been provided for kink collisions since, up
to our knowledge, they have not been considered before for Nambu-Goto
strings. Their importance has however been discussed for superstrings
having Y-junctions in Refs.~\cite{Binetruy:2010cc, Bohe:2011rk}. In
section~\ref{sec:spec}, we present our results, namely the stochastic
gravitational wave spectra expected when loops contain either one
cusp, or one kink, or two kinks plus one collision per
oscillation. Our findings are critically compared to previous results
and when we do not include the above-mentioned new effects, our
prediction are in very good agreement with the one of
Ref.~\cite{Blanco-Pillado:2013qja} for $\GU=10^{-7}$. For lower values
of $\GU$, small differences in the power law exponents of the loop
distribution, as well as relaxation effects encoded only within our
Boltzmann approach, start to play a role and this agreement is
accordingly degraded. Physical explanations of the new features are
provided along with simple analytic estimates. Finally, in
section~\ref{sec:micro}, we explore various motivated models for the
string microstructure and estimate how the string tension $\GU$ is
constrained by the current Laser Interferometer Gravitational-Wave
Observatory (LIGO) and European Pulsar Timing Array (EPTA)
observations.

\section{Gravitational wave sources}
\label{sec:src}

\subsection{Basic assumptions}

The loops we are interested in are of sub-Hubble sizes such that one
can assume their dynamics over one oscillation to be approximately
described by the Nambu-Goto equations of motion in Minkowski
spacetime. Indeed, as can be seen in figure~\ref{fig:ldtrans}, the
distribution vanishes on Hubble length scales, i.e., for $\gamma \ge
\gammai$. Let us stress that, in addition to the loops discussed here,
a scaling network of cosmic strings exhibit a few large loops with
sizes typical of the long strings correlation length, referred to as
Kibble loops (see Ref.~\cite{Ringeval:2010ca}). These loops are in all
points similar to the infinite strings, and so is their contribution
to the stochastic gravitational waves
background~\cite{Sakellariadou:1990ne}. For convenience, we will take
$\gammai=1$ in the following, both in the matter and radiation eras. As
discussed in Ref.~\cite{Damour:2001bk}, the size of the loops allows a
length scale separation in the derivation of the GW signal. In a local
wave zone around each loop, GW emission can be estimated by
linearising the metric around a Minkowski background $g_{\mu\nu} =
\eta_{\mu\nu} + h_{\mu\nu}$. The emitted GWs then propagate from the
local wave zone to the observer along light-like geodesics of an
assumed flat Friedman-Lema\^{\i}tre-Robertson-Walker (FLRW) metric
\begin{equation}
\ud s^2 = a^2(\eta) \left(-\ud \eta^2 + \ud \chi^2 + \chi^2 \ud
\Omega^2 \right).
\end{equation}
Here $\eta$ and $\chi$ are the conformal time and conformal radius of
a spherical coordinate system centered on the source, and $a(\eta)$ is the
scale factor. For wavelengths much smaller than the Hubble radius, in
addition to frequency redshifting, one finds that the amplitude of
$a(\eta) \chi h_{\mu\nu}(\eta,\chi)$ is conserved during propagation
thereby allowing to relate the observed GW strain to the one derived
around each loop~\cite{Damour:2001bk}.

In practice, defining
\begin{equation}
  \hit_{\mu \nu} \equiv h_{\mu \nu} - \dfrac{1}{2} h \eta_{\mu\nu},
  \qquad h \equiv h_{\mu}^{\mu},
\end{equation}
supplemented with the harmonic gauge condition $\hit^{\mu
  \nu}_{~,\nu}=0$, the linearised Einstein equations are solved by a
retarded propagator and one gets
\begin{equation}
\hitlwz^{\mu\nu}(r) = 4 G \int\ud t' \ud^3 \br' \dfrac{T^{\mu
    \nu}(r')}{|\br - \br'|} \dirac{t-t' - |\br-\br'|},
\label{eq:Gret}
\end{equation}
where $G$ is the Newton constant. In this expression $r$ stands for
the four-vector at which one evaluates the GW strain while $t$
and $\br$ denotes its time and spatial components, respectively. The
index ``$\flat$'' is a reminder that the result holds only in the
local wave zone assuming Minkowski background. Both the stess tensor
$T^{\mu\nu}$ and the equations of motion for a cosmic string loop can
be obtained from the Nambu-Goto action
\begin{equation}
S = -U \int \ud \tau \ud \sigma \sqrt{(\Xd \Xp)^2 - \Xd^2 \Xp^2}\,.
\label{eq:Sng}
\end{equation}
The quantities $X^\mu(\tau,\sigma)$ are the embedding functions of the
string worlsheet, $\tau$ and $\sigma$ stand for the time-like and
space-like internal coordinates, $\Xp^\mu \equiv \partial
X^\mu/\partial \sigma$ and $\Xd^\mu \equiv \partial X^\mu/\partial
\tau$ (the square root is the determinant of the induced metric along
the string worldsheet). From Eq.~\eqref{eq:Sng}, in the temporal and
transverse gauge where $X^0 = \tau = \eta$ and $\bXp\cdot \bXd = 0$,
the stress tensor reads
\begin{equation}
T^{\mu \nu}(r) = U \int \ud \sigma \left(-\Xp^\mu \Xp^\nu +
  \Xd^\mu \Xd^\nu \right) \diracb{\br-\bX(\tau,\sigma)},
\label{eq:NGstress}
\end{equation}
while the string dynamics is given by the propagation of left and
right moving string deformations~\cite{Vilenkin:2000}
\begin{equation}
\bX(\tau,\sigma) = \dfrac{1}{2} \left[ \bX_+(\sigma_+) +
  \bX_-(\sigma_-) \right].
\label{eq:eom}
\end{equation}
The vectors $\bX_\pm$ are constant along the characteristics
$\sigma_\pm \equiv \sigma \pm \tau$ and verify $|\bXp_\pm|^2=1$. By
definition of a loop, periodicity conditions in its rest frame require
$\bX(\tau, \sigma + \ell) = \bX(\tau, \sigma)$ such that
$\bX_\pm(\sigma_\pm + \ell)=\bX_\pm(\sigma_\pm)$ are also of period
$\ell$. From Eq.~\eqref{eq:eom} together with Lorentz invariance along
the worldsheet imply that loops oscillate with a period $T=\ell/2$ as
$\bX(\tau + \ell/2,\sigma+\ell/2) = \bX(\tau,\sigma)$. As a result, in
Fourier space, one has
\begin{equation}
T^{\mu \nu}(\varpi_n,\bkappa) = \dfrac{1}{T} \int \ud^3 \bx \int_0^{T} \ud
t \, T^{\mu \nu}(x) \, e^{i \varpi_n t - i\bkappa \cdot \bx},
\label{eq:FTconv}
\end{equation}
where $n$ is an integer and
\begin{equation}
\varpi_n = \dfrac{2 \pi n}{T} = \dfrac{4 \pi n}{\ell}\,.
\end{equation}
Plugging Eqs.~\eqref{eq:NGstress} and \eqref{eq:eom} into
Eq.~\eqref{eq:FTconv} allows us to obtain an explicit expression for the
time domain Fourier transform of the GW tensor in
Eq.~\eqref{eq:Gret}. After some algebra, one gets~\cite{Damour:2001bk}
\begin{equation}
\hitlwz^{\mu\nu}(\varpi_n,|\br|\bnhat) = \dfrac{G U}{T} \dfrac{e^{i \varpi_n
    |\br|}}{|\br|} C^{\mu \nu}, \qquad C^{\mu \nu}\ \equiv I_+^\mu I_-^\nu + I_+^\nu I_-^\mu,
\label{eq:hitlwz}
\end{equation}
where quadratic terms $\order{1/|\br|^2}$ have been neglected far from
the source. In this expression, the integrals $I_\epsilon^\mu$ (with
$\epsilon=\pm$) stand for
\begin{equation}
  I_\epsilon^\mu(\varpi_n,\bkappa) \equiv \int \ud \sigma_\epsilon
  \exp\left(\dfrac{i \varpi_n \sigma_\epsilon}{2} - \dfrac{i
  \bkappa \cdot \bX_\epsilon}{2}\right) \dfrac{\ud
    X_\epsilon^\mu}{\ud \sigma_\epsilon},
\label{eq:Ipm}
\end{equation}
and should be evaluated on the light cone, namely for $\bkappa = \varpi_n
\bnhat$, $\bnhat$ being an unit vector in the direction of
propagation, i.e. $\kappa_n \equiv (\varpi_n,\varpi_n\bnhat)$ is
light-like. Notice that $X_\epsilon \equiv
(\sigma_\epsilon,\bX_\epsilon)$ and its derivative $\Xp_\epsilon^\mu
\equiv \ud X^\mu_\epsilon/\ud \sigma_\epsilon$ is also light-like.

Embedding the local wave zone into FLRW spacetime, one can identify
$r= \aem \chi$, $\chi$ being the comoving coordinate of a FLRW
coordinate system centered on the source and $\aem$ the scale factor
at the time of emission. Including redshifting and using the
conservation of $a \chi h^{\mu\nu}$ during propagation, one finally
gets for the observed GW strain
\begin{equation}
\hit^{\mu \nu}(\etaobs,\chiobs) = \dfrac{G U}{T \chiobs}
\sum_{n=-\infty}^{+\infty} e^{i \aem \varpi_n \left(\chiobs -
  \etaobs \right)} C^{\mu\nu}(\kappa_n).
\label{eq:hitobs}
\end{equation}
Notice that in a coordinate system centered on the observer, $\chiobs$
is also the comoving distance to the source $\chiobs = \chiem$

\subsection{Cusps, kinks and collisions}
\label{sec:ckandkk}
The amplitude tensor $C^{\mu \nu}$ being the product of rapidly
oscillating integrals $I^\mu_\epsilon$, one can qualitatively infer
the large frequency behaviour of the GW strain. From
Eq.~\eqref{eq:Ipm}, by the Riemann-Lebesgue lemma, the
$I^\mu_\epsilon$ are expected to always decay exponentially fast with
$\varpi_n$ unless the phase $\varphi=\varpi_n \sigma_\epsilon -
\bkappa \cdot \bX_\epsilon$ has a saddle point, and/or the function
$\Xp^\mu_\epsilon$ is not smooth.

Saddle points for $I^\mu_\epsilon(\kappa_n)$ occur when $\bnhat \cdot
\bXp_\epsilon = 1$, i.e. when $\bXp_\epsilon$ is aligned with the
direction of propagation $\bnhat$. Requiring both $I^\mu_+$ and
$I^\nu_-$ to have a saddle point therefore implies that $\bnhat =
\bXp_+ = \bXp_-$, the last equality being the condition for a cosmic
string loop to form a ``cusp''. In terms of worldsheet coordinates,
denoting by $\ssaddle_\epsilon$ the values at which the vectors
$\bXp_\epsilon$ coincide with $\bnhat$, the cusp appears at a specific
instant and location given by $\tsaddle = (\ssaddle_+ - \ssaddle_-)/2$
and $\ssaddle= (\ssaddle_+ + \ssaddle_-)/2$. At this point, $\bXd^2=1$
and the cusp emits a beamed GW burst along the direction $\bXp_+ =
\bXp_-$. The beam opening angle has been derived in
Ref.~\cite{Damour:2001bk} and reads
\begin{equation}
\thetab = \left(\dfrac{8 \pi}{\sqrt{3} \varpi \ell} \right)^{1/3},
\label{eq:thetab}
\end{equation}
where we have omitted the subscript $n$ for $\varpi$ to indicate that 
the frequency can be considered almost continuous in the large $n$ limit
we are interested in.

Discontinuities in the functions $\Xp^\mu_\epsilon(\sigma_\epsilon)$
are common features for cosmic strings and correspond to kinks in the
shape of the string. As a result, another configuration for which $C^{\mu
  \nu}$ does not decrease exponentially fast with frequencies is when
one of the integral $I^\mu_\epsilon$ develops a saddle point at
$\ssaddle_\epsilon$ and the other $I^\nu_{-\epsilon}$ a kink at
$\skink_{-\epsilon}$. From Eq.~\eqref{eq:eom}, the string worldsheet
$\bX$ has indeed a kink propagating along one direction $\sigma =
\skink_{-\epsilon} + \epsilon \tau$. Along these locations,
the saddle point of $I^\mu_\epsilon$ occurs along the $\bXp_\epsilon$
direction only such that GW are beamed in a light-house manner by the
propagating kink.

Both of these situations have been proposed and exhaustively studied
in Ref.~\cite{Damour:2001bk} and we simply quote the result. Keeping
the dominant terms of a Taylor expansion around the saddle point at
$\sigma_\epsilon = \ssaddle_\epsilon$, one gets for the cusp case
\begin{equation}
\left. I^\mu_\epsilon(\kappa) \right|_{\uc} \simeq i
\dfrac{\varpi}{|\varpi|} \left(\dfrac{2}{3}\right)^{1/3} \dfrac{4
  \pi}{\Gamma\left(\dfrac{1}{3}\right)}
\dfrac{\Xpp_\epsilon^\mu\left(\ssaddle_\epsilon\right)}{\left|
  \bXpp_\epsilon\left(\ssaddle_\epsilon\right)\right|^{4/3}
  |\varpi|^{2/3}}\,.
\label{eq:Icusp}
\end{equation}
From dimensional arguments, one can take $|\bXpp_\epsilon| = 2 \pi
\beta/ \ell$ with $\beta = \order{1}$ and the overall numerical factor
\begin{equation}
\etas \equiv \left(\dfrac{2}{3}\right)^{1/3} \dfrac{4\pi}{\Gamma\left(\dfrac{1}{3}\right)}\,,
\label{eq:etas}
\end{equation}
is approximately $\etas \simeq 4.1$. For kinks, one can define the
discontinuity amplitude vectors $u^\mu_{\epsilon 1}$ and
$u^\mu_{\epsilon 2}$ such that
\begin{equation}
\lim_{\sigma_\epsilon \rightarrow \skink_\epsilon - 0}
\Xp^\mu_\epsilon(\sigma_\epsilon) = u^\mu_{\epsilon 1}, \qquad
\lim_{\sigma_\epsilon \rightarrow \skink_\epsilon + 0}
\Xp^\mu_\epsilon(\sigma_\epsilon) = u^\mu_{\epsilon 2}.
\end{equation}
Both $u_{\epsilon 1}$ and $u_{\epsilon 2}$ are null vectors. At
leading order, the integrals \eqref{eq:Ipm} read
\begin{equation}
\left. I^\mu_\epsilon(\kappa) \right|_{\uk} \simeq 2i
\left(\dfrac{u^\mu_{\epsilon 1}}{\kappa \cdot u_{\epsilon 1}} -
    \dfrac{u^\mu_{\epsilon 2}}{\kappa \cdot u_{\epsilon 2}} \right) =
    \dfrac{2i}{\varpi} v^\mu_\epsilon,
\label{eq:Ikink}
\end{equation}
where we have defined
\begin{equation}
v^\mu_\epsilon \equiv \dfrac{u^\mu_{\epsilon 1}}{\hat{\kappa} \cdot
  u_{\epsilon 1}} - \dfrac{u^\mu_{\epsilon 2}}{\hat{\kappa} \cdot
  u_{\epsilon 2}}\,, \qquad \hat{\kappa} \equiv (1,\bnhat). \label{def-v}
\end{equation}
From Eqs.~\eqref{eq:Icusp} and \eqref{eq:Ikink}, one concludes that,
at high frequencies, $C^{\mu\nu} \propto \varpi^{-4/3}$ for cusps and
$C^{\mu\nu} \propto \varpi^{-5/3}$ for kinks. As a result, if cusps
develop on a loop, they should dominate GW
emission~\cite{Damour:2001bk}.

However, such a conclusion may not hold if the number of kinks
propagating on a cosmic string loop is large, a situation which may
occur even in presence of GW backreaction~\cite{Wachter:2016rwc,
  Wachter:2016hgi}. As a matter of fact, kinks themselves render the
appearance of cusps less
likely~\cite{Blanco-Pillado:2015ana}. Moreover, because kinks are
formed during string intersections, they are created in pairs and each
kinky loop is expected to have an equal number of left- and
right-moving kinks. This suggests that, in addition to the two
previously discussed situations, another significant source of GW
emission occurs when both integrals $I^\mu_\epsilon$ and
$I^\nu_{-\epsilon}$ behave as in Eq.~\eqref{eq:Ikink}, which implies
that $C^{\mu\nu} \propto \varpi^{-2}$ at large frequencies. Fixing
both $\sigma_+ = \skink_+$ and $\sigma_- = \skink_-$ corresponds to an
unique event along the worldsheet, at $\tkink = (\skink_+ -
\skink_-)/2$ and $\skink = (\skink_+ + \skink_-)/2$, which describes
the collision of two kinks moving in opposite directions. The GW
emission does not depend on $\bnhat$ and is therefore
isotropic. Although $C^{\mu \nu}$ is decreasing faster with
frequencies than in the cusp and kink case, one should keep in mind
that the number of kink collisions per loop oscillation scales as the
square of the number of kinks. Indeed, if a loop exhibits $\Nkink$
kinks, in an equal number of left-movers and right-movers, the number
of collisions per loop oscillation is $\Nkink^2/4$. As we show later
on, kinks collision may actually dominate the overall GW signal for
very kinky loops.

\subsection{Single loop energy and spectral density}
\label{sec:rhogw}

From the pseudo stress tensor associated with $h_{\mu\nu}$, one can
estimate the mean energy density associated with the waveform of
Eq.~\eqref{eq:hitobs} at the observer's location. Taking the scale
factor today $a_0=1$ and averaging over the observed loop oscillation
period, i.e. $T/\aem$, one gets
\begin{equation}
\rhogw^{(\circ)}= \dfrac{1}{32 \pi G} \dfrac{\aem}{T} \int_0^{T/\aem}
\left(\dot{\hit}_{\alpha \beta} \dot{\hit}^{\alpha \beta} -
\dfrac{1}{2} \dot{\hit}^2 \right) \ud \eta,
\end{equation}
where a dot denotes derivative with respect to the conformal
time. Plugging Eq.~\eqref{eq:hitobs} into this expression yields
\begin{equation}
\rhogw^{(\circ)} = \dfrac{1}{16 \pi G} \left(\dfrac{\GU}{T
  \chiem}\right)^2 \, \sum_{n=1}^{+\infty} \left(\aem \varpi_n \right)^2
\left[ C^*_{\alpha \beta}(\kappa_n) C^{\alpha \beta}(\kappa_n) -
  \dfrac{1}{2} \left|C(\kappa_n) \right|^2 \right].
\label{eq:rhogws}
\end{equation}
Each angular frequency $\varpi_n$ is redshifted and observed as $\omega_n
\equiv \aem \varpi_n$. Moreover, Eq.~\eqref{eq:rhogws} can be
further reduced by taking the continuum limit over $\omega_n$
\begin{equation}
\rhogw^{(\circ)} = \dfrac{T}{32 \pi^2 G \aem} \left(\dfrac{\GU}{T
  \chiem} \right)^2 \int_{\omega_1}^{+\infty} \ud \omega \, \omega^2 \left[
  C^*_{\alpha \beta}\left(\dfrac{k}{\aem}\right) C^{\alpha
    \beta}\left(\dfrac{k}{\aem}\right) - \dfrac{1}{2} \left|C\left(\dfrac{k}{\aem}\right)
  \right|^2 \right],
\label{eq:rhogwscont}
\end{equation}
where the four-vector $k\equiv (\omega,\omega \bnhat)$. Notice that
the integral has a lower cut-off in frequencies which corresponds to
the fundamental mode $n=1$, i.e.
\begin{equation}
\omega_1(\ell,\zem) = \dfrac{2 \pi \aem}{T} = \dfrac{4 \pi}{(1+\zem)\ell}\,.
\label{eq:omega1}
\end{equation}

We define the spectral density parameter of GW today,
$\OmegaGW(\omega)$, by
\begin{equation}
\dfrac{\rhogw}{\rhocrit} \equiv \int_0^{+\infty} \dfrac{\ud
  \omega}{\omega} \OmegaGW(\omega)\,,
\end{equation}
where the critical density is $\rhocrit=3 \Ho^2/(8 \pi G)$, $\Ho$
being the Hubble parameter today. Comparing this expression to
Eq.~\eqref{eq:rhogwscont}, we obtain for a single loop, assuming the
observer to be within the beam
\begin{equation}
\OmegaGW^{(\circ)}(\omega) = \dfrac{(\GU)^2}{12 \pi \Ho^2} \, \dfrac{\omega^3}{T
  \aem \chiem^2} \left[
  C^*_{\alpha \beta}\left(\dfrac{k}{\aem}\right) C^{\alpha
    \beta}\left(\dfrac{k}{\aem}\right) - \dfrac{1}{2} \left|C\left(\dfrac{k}{\aem}\right)
  \right|^2 \right] \heaviside{\omega - \omega_1}.
\label{eq:OmegaGWsgle}
\end{equation}
In order to estimate the observed stochastic GW spectrum, one has
still to define which events are stochastic from the observer point of
view and add up the contribution of all loops in the Universe.

\section{Stochastic gravitational wave spectrum}
\label{sec:spec} 

From the cosmological loop distribution $\calF(\ell,z)$, the number of
loops of size $\ell$ at redshift $z$ per $\ud\ell \ud z$ volume in a
flat FLRW metric reads
\begin{equation}
\dfrac{\ud N_\circ}{\ud \ell \ud z} = \dfrac{\ud V}{\ud
  z}\calF(\ell,z), \qquad \dfrac{\ud V}{\ud z} =  \dfrac{4 \pi \chi^2(z)}{{(1+z)}^3 H(z)}\,,
\end{equation}
where $H(z)$ stands for the Hubble parameter at redshift $z$. Assuming
for the time being that all of these loops have only one GW emission
event per oscillation, either a cusp, a kink or a collision, one can
estimate their contribution to the GW event rate today, at angular frequency
$\omega$, by
\begin{equation}
\dfrac{\ud N_\ugw }{\ud z \ud \ell \ud \eta} = \dfrac{2}{(1+z)\ell}
\dfrac{\ud V}{\ud z} \calF(\ell,z) \heavisideb{\omega -
  \omega_1(\ell,z)}\Deltab(\omega,\ell,z),
\label{eq:rateall}
\end{equation}
where the time averaging is performed over one oscillation period
today, i.e. for $\Delta \eta_0 = (1+z)T$. This only makes sense for
frequencies higher than the redshifted fundamental mode $\omega_1$, as
it was the case for Eq.~\eqref{eq:rhogwscont}. The beam factor
$\Deltab(\omega,\ell,z)=\Omegab/(4\pi)$ is the probability for the
observer to be within the GW emission solid angle $\Omegab$. From
Eq.~\eqref{eq:thetab}, it reads
\begin{equation}
\begin{aligned}
  \left. \Deltab(\omega,\ell,z) \right|_{\uc} & = \dfrac{\pi \thetab^2}{4
  \pi} =  \dfrac{1}{4} \left[\dfrac{8 \pi}{\sqrt{3} \omega (1+z)
      \ell}\right]^{2/3}, \\
  \left. \Deltab(\omega,\ell,z) \right|_{\uk} & = \dfrac{2 \pi \thetab}{4
  \pi} = \dfrac{1}{2} \left[\dfrac{8 \pi}{\sqrt{3} \omega (1+z) \ell} \right]^{1/3},
\end{aligned}
\label{eq:beams}
\end{equation}
for cusp and kink, respectively, while $\Deltab(\omega,\ell,z)|_{\ukk}
= 1$ for kinks collision.

Even for a perfect GW detector with infinite sensitivity, all of these
events superimpose and may not be detected individually, especially if
their rate of occurrence exceeds the detector frequency, here assumed
to be $f=\omega/(2\pi)$. As pointed in Ref.~\cite{Damour:2001bk}, all
individually separable events will be observed as bursts and cannot be
counted within the stochastic background. In order to separate the
burst-like events from the stochastic ones, we have followed the
method of Ref.~\cite{Siemens:2006yp, Olmez:2010bi}. It assumes that
the event rate is a monotonic decreasing function of the GW strain
power such that the separation between burst-like and stochastic-like
event can be made on the values of $|h_{\mu \nu}(\omega)|^2$. Because
we do not attempt to disambiguate the polarization degrees of freedom,
let us define
\begin{equation}
\hitrms^2(\etaobs,\chiobs) \equiv \left \langle \hit_{\mu\nu}^*
\hit^{\mu\nu} \right \rangle -
\dfrac{1}{2} \left \langle \left|\hit\right|^2 \right \rangle,
\end{equation}
where $\langle A \rangle$ is average of quantity $A$ over a redshifted
oscillation period. From Eq.~\eqref{eq:hitobs}, averaging over a
redshifted oscillation period $\Delta \eta_0$ and taking the continuum
limit in $\varpi_n$, one gets per event:
\begin{equation}
\hitrms^2(\etaobs,\chiobs) = \dfrac{2T}{\aem} \left(\dfrac{\GU}{T
  \chiobs} \right)^2 \int_{\omega_1}^{+\infty} \dfrac{\ud \omega}{2
  \pi} \left[ C^*_{\alpha \beta}\left(\dfrac{k}{\aem}\right) C^{\alpha
    \beta}\left(\dfrac{k}{\aem}\right) - \dfrac{1}{2}
  \left|C\left(\dfrac{k}{\aem}\right) \right|^2 \right].
\end{equation}
The GW strain power at frequency $\omega$ per event is thus given by
\begin{equation}
\hitrms^2(\omega,\ell,z)  = \left[\dfrac{\GU(1+z)}{\chi(z)}\right]^2
\Cbar^2\left[k(1+z)\right] \heavisideb{\omega - \omega_1(\ell,z)},
\label{eq:hitrms}
\end{equation}
where we have defined
\begin{equation}
\Cbar^2 \equiv C^*_{\alpha \beta}C^{\alpha \beta} - \dfrac{1}{2} \left|C \right|^2.
\label{eq:Cbar}
\end{equation}
From Eq.~\eqref{eq:rateall}, the minimal amplitude $\hitstar(\omega)$
under which burst-like events will not be time separable any more is
solution of
\begin{equation}
\iint \ud z \ud \ell \dfrac{2}{(1+z)\ell} \dfrac{\ud V}{\ud z}
\calF(\ell,z) \Deltab(\omega,\ell,z) \heavisideb{\omega -
  \omega_1(\ell,z)}
\heavisideb{\hitrms(\omega,\ell,z)-\hitstar(\omega)} =
\dfrac{\omega}{2\pi}\,.
\label{eq:hitstar}
\end{equation}

From Eqs.~\eqref{eq:OmegaGWsgle}, \eqref{eq:rateall},
\eqref{eq:hitrms} and \eqref{eq:hitstar}, one obtains for the total
stochastic gravitational wave spectrum
\begin{equation}
\begin{aligned}
  \OmegaSGW(\omega) & = \dfrac{(\GU)^2 \omega^3 }{6 \pi \Ho^2} \iint \ud z
\ud \ell \dfrac{\ud V}{\ud z} \dfrac{\calF(\ell,z)}{\ell(1+z)}
\Deltab(\omega,\ell,z) \\ &\times \dfrac{(1+z)^2}{\chi^2(z)}
\Cbar^2(\omega,\ell,z) \heavisideb{\omega - \omega_1(\ell,z)}
\heavisideb{\Cbarstar(\omega,z) - \Cbar(\omega,\ell,z)},
\label{eq:omsgwlt}
\end{aligned}
\end{equation}
where the rescaled stochastic threshold $\Cbarstar$ stands for
\begin{equation}
\Cbarstar(\omega,z) \equiv \dfrac{\chi(z)}{(1+z)\GU} \hitstar(\omega).
\end{equation}
Notice that the previous expressions are valid assuming only one cusp,
or one kink, or one collision event per loop oscillation. The
generalization to multiple events of one kind is however trivial as it
is enough to rescale, for instance, $\calF$ by the number of events
per loop oscillation in all the previous formulas.

\subsection{Unified source functions and scaling variables}
\label{sec:scalvars}

As discussed in the introduction, up to the transition from radiation
to matter and the late-time acceleration of the Universe, the loop
scaling distribution is stationary in terms of the variable $\gamma
=\ell/t$, where $t$ is the cosmic time. It is therefore convenient to
change variables from $(\ell,z)$ to $(\gamma,z)$ in
Eq.~\eqref{eq:omsgwlt}. After expanding the various cosmological
factors, one gets
\begin{equation}
\begin{aligned}
  \OmegaSGW(\omega) & = \dfrac{2 (\GU)^2 \omega^3}{3 \Ho^3} \iint \ud z
\ud \gamma \dfrac{t^4(z) \calF(\gamma,z)}{\gamma (1+z)^2} \dfrac{\Ho}{H(z)}
\Deltab(\omega,\gamma,z) \\
& \times \Ctilde^2(\omega,\gamma,z) \, \heavisideb{\omega - \dfrac{4
    \pi}{t(z) (1+z) \gamma}} \heavisideb{\Ctildestar(\omega,z)
  - \Ctilde(\omega,\gamma,z)},
\end{aligned}
\label{eq:onesgwgz}
\end{equation}
in which we have rendered explicit the stationary scaling
distribution $t^4 \calF$. We have also introduced dimensionless
quantities such as $\Ctilde^2 \equiv \Cbar^2 / t^4(z)$. Accordingly,
one can define the dimensionless stochastic threshold
\begin{equation}
\hittstar(\omega) \equiv \dfrac{\Ho}{\GU} \hitstar(\omega),
\end{equation}
such that
\begin{equation}
\Ctildestar(\omega,z) = \dfrac{\chi(z)}{\Ho t^2(z) (1+z)} \hittstar(\omega),
\end{equation}
is also dimensionless. From Eq.~\eqref{eq:hitstar}, in terms of
$\gamma$ and $z$, $\hittstar(\omega)$ is obtained by solving the rate
equation
\begin{equation}
\begin{aligned}
  & \iint \ud z \ud \gamma \dfrac{t^4(z) \calF(\gamma,z)}{\gamma
    (1+z)^4} \dfrac{\Ho}{H(z)} \dfrac{\chi^2(z)}{t^2(z)}
  \dfrac{8\pi \Deltab(\omega,\gamma,z)}{t^2(z) \Ho^2} \\ & \times
  \heavisideb{\omega - \dfrac{4 \pi}{t(z) (1+z) \gamma}}
  \heavisideb{\Ctilde(\omega,\gamma,z) - \dfrac{\chi(z)}{\Ho
      t^2(z)(1+z)} \hittstar(\omega)} = \dfrac{\omega}{2 \pi \Ho}\,.
\end{aligned}
\label{eq:burstrategz}
\end{equation}
Depending on the GW source type, cusp, kink or collision, both
$\Deltab$ and $\Ctilde$ are changed such that these two equations
should be solved separately for each case. However, from
Eq.~\eqref{eq:beams}, one can see that all beam factors are of the
form
\begin{equation}
\Deltab(\omega,\gamma,z) = \dfrac{b_\alpha}{\gamma^{1-1/\alpha}} \left[ \dfrac{8 \pi}{\sqrt{3}
    \omega t(z) (1+z)} \right]^{1-1/\alpha}\,,
\label{eq:beam}
\end{equation}
where the parameter $\alpha=3$ for cusps, $\alpha=3/2$ for kinks and
$\alpha=1$ for collisions. The coefficient $b_\alpha$ is a numerical
constant equals to $1/4$, $1/2$ or $1$, respectively. Plugging the
integrals $I^\mu_\epsilon$ quoted in Eqs.~\eqref{eq:Icusp} and
\eqref{eq:Ikink} into Eq.~\eqref{eq:Cbar}, one gets
\begin{equation}
\begin{aligned}
  \Ctilde_{\uc}(\omega,\gamma,z) & = \dfrac{\etas^2 \sqrt{2} }{(2 \pi \beta)^{2/3}}
\, \dfrac{\gamma^{2/3}}{\left[\omega t(z) (1+z) \right]^{4/3}}\,, \\
\Ctilde_{\uk}(\omega,\gamma,z) &= \dfrac{2\etas \sqrt{2 v_\pm^2}}{(2\pi\beta)^{1/3}}
\, \dfrac{\gamma^{1/3}}{\left[\omega t(z) (1+z)\right]^{5/3}}\,, \\
\Ctilde_{\ukk}(\omega,\gamma,z) &= 4 \sqrt{2 v_+^2 v_-^2} \,
\dfrac{1}{\left[\omega t(z) (1+z)\right]^2}\,,
\end{aligned}
\label{eq:Ctildes}
\end{equation}
for the cusp, kink and collision case. Let us notice that $u_{\epsilon
  1}$, $u_{\epsilon 2}$ and ${\hat \kappa}$ in Eq.~(\ref{def-v}) are
all future directed null vectors and this ensures that both $v_+^2$
and $v_-^2$ are positive. As a result, all the source functions
$\Ctilde(\omega,\gamma,z)$ can also be unified as
\begin{equation}
\Ctilde(\omega,\gamma,z) = c_\alpha \dfrac{\gamma^{1-1/\alpha}}{\left[\omega t(z) (1+z)\right]^{1+1/\alpha}}\,.
\end{equation}
The numerical constants $c_\alpha$ can be read from
Eq.~\eqref{eq:Ctildes} and contains all the theoretical uncertainties
associated with each of the GW source type
\begin{equation}
c_{3} \equiv \dfrac{\etas^2 \sqrt{2} }{(2 \pi \beta)^{2/3}}\,, \qquad
c_{3/2} \equiv \dfrac{2\etas \sqrt{2 v_\pm^2}}{(2\pi\beta)^{1/3}}\,, \qquad
c_{1} \equiv  4 \sqrt{2 v_+^2 v_-^2}\,.
\label{eq:calpha}
\end{equation}

From Eqs.~\eqref{eq:onesgwgz} and \eqref{eq:burstrategz}, these
theoretical uncertainties can be trivially absorbed into our final set
of rescaled dimensionless variables defined by
\begin{equation}
  \Chat(\omega,\gamma,z) \equiv
  \dfrac{\Ctilde(\omega,\gamma,z)}{c_\alpha}\,, \qquad
  \hithstar(\omega) \equiv \dfrac{\hittstar(\omega)}{c_\alpha} =
  \dfrac{\Ho}{\GU c_\alpha} \hitstar(\omega), \qquad
  \OmegaSGWH(\omega) \equiv \dfrac{\OmegaSGW(\omega)}{c_\alpha^2}\,.
\label{eq:hatsdef}
\end{equation}
All ``hat'' quantities are therefore independent of the value of
$\etas$ and $\beta$ while the physical quantities, as for instance
$\OmegaSGW(\omega)$, can be recovered from the above formulas.

\subsection{Loop visibility domains}
\label{sec:domains}

The Heaviside functions in Eqs.~\eqref{eq:onesgwgz} and
\eqref{eq:burstrategz} directly stem from the existence of the minimal
frequency $\omega_1$ and the requirement of keeping only
stochastic-like events. In terms of the variables $(\gamma,z)$, they
define some integration domains which physically take into account
frequency redshifting as well as the fact that a GW burst-like signal
may either come from a very distant high amplitude emission event or
a close small amplitude one. Let us define the function $\gammaw(\omega,z)$
by
\begin{equation}
\gammaw(\omega,z) \equiv \dfrac{4 \pi}{\Ho t(z)(1+z)} \dfrac{\Ho}{\omega}\,.
\label{eq:gammaone}
\end{equation}
From this definition, the first Heaviside function in both
Eqs.~\eqref{eq:onesgwgz} and \eqref{eq:burstrategz} will be
non-vanishing only for $\gamma \ge \gammaw(\omega,z)$. The fundamental
mode of each loop therefore imposes a lower cut-off in the size of the
loops that are ``visible'' at the angular frequency $\omega$.

Similarly, we can define, for $\alpha \ne 1$, the length scale
\begin{equation}
\gammas(\omega,z,\hithstar) \equiv \left[\hithstar
  \dfrac{\chi(z)}{t(z)} \right]^{\frac{\alpha}{\alpha-1}} \left[\Ho t(z)
  (1+z)\right]^{\frac{1}{\alpha-1}} \left(\dfrac{\omega}{\Ho}\right)^{\frac{\alpha+1}{\alpha-1}},
\label{eq:gammastar}
\end{equation}
such that events counted in the burst rate only come from the domain
$\gamma \ge \gammas$, see the second Heaviside function in
Eq.~\eqref{eq:burstrategz}. Conversely, stochastic events appearing in
Eq.~\eqref{eq:onesgwgz} are those verifying $\gamma < \gammas$. Kink
collision, $\alpha=1$, is a particular case and will be discussed
below.

\begin{figure}
  \begin{center}
    \includegraphics[width=\onefigw]{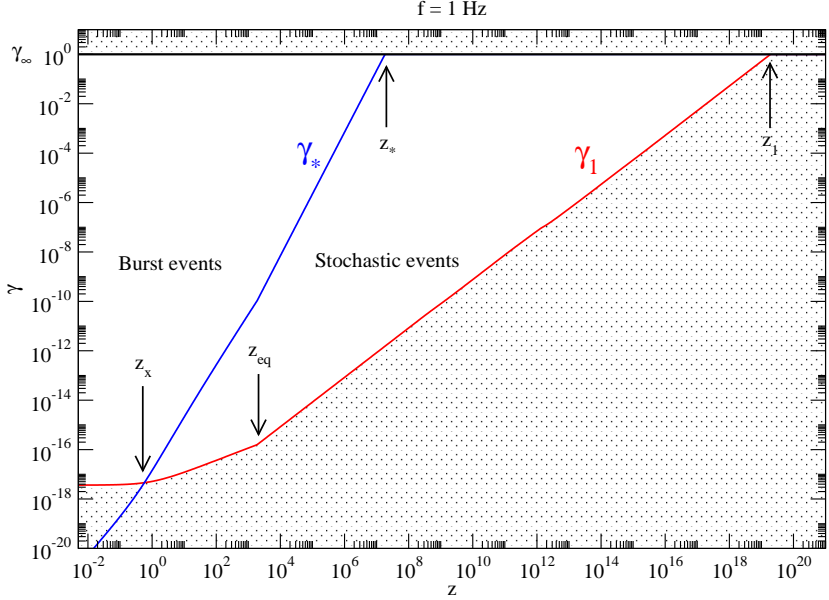}
    \caption{Loop visibility domains for GW kink events observed at
      frequency $f=1\,\Hz$ in the plane $(\gamma=\ell/t,z)$. Only
      loops having $\gamma > \gamma_1(\omega,z)$ oscillate slow enough
      to have a redshifted fundamental mode $\omega_1 < \omega$. The
      curve $\gammas(\omega,z,\hithstar)$ separates the domain in two
      regions. For $\gamma < \gammas$, the observed GW events occur
      fast enough to be interpreted as stochastic, whereas they could
      be individually isolated for all loops having $\gamma >
      \gammas$. For the illustration, we have chosen $\hithstar \simeq
      5\times 10^{-37}$, which solves the rate
      equation~\eqref{eq:burstrategz} for a loop distribution typical
      of $\GU=10^{-7}$.}
    \label{fig:domains_1Hz}
  \end{center}

\end{figure}
\begin{figure}
  \begin{center}
    \includegraphics[width=\onefigw]{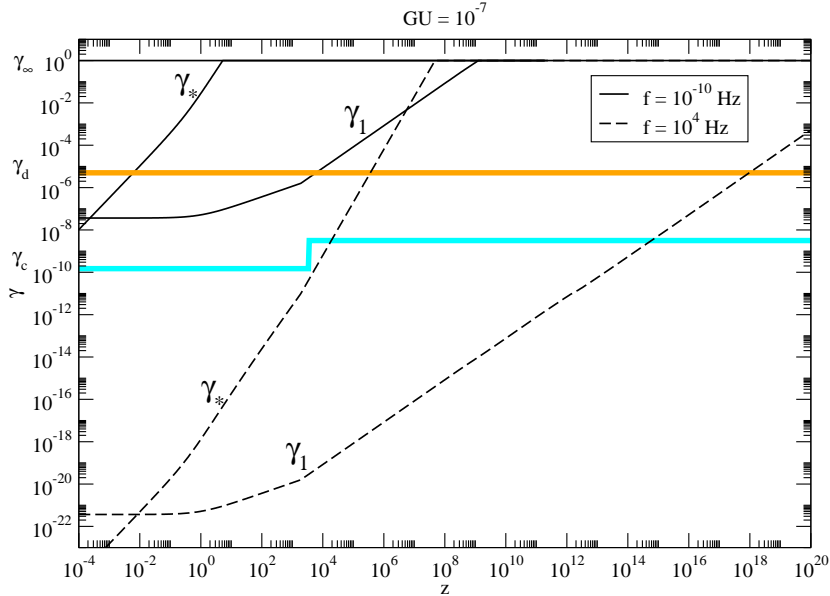}
    \caption{Dependence of the burst and stochastic domains in the
      plane $(\gamma,z)$ with respect to the observation frequency
      $f=\omega/(2\pi)$ (cusp case with $\alpha=3$). The solid lines
      represent the boundary functions $\gammaw(\omega,z)$ and
      $\gammas(\omega,z,\hithstar)$ for $f=10^{-10}\,\Hz$ while the
      dashed lines are for $f=10^4\,\Hz$. Values for
      $\hithstar(\omega)$ have been chosen to solve the rate equation
      for $\GU = 10^{-7}$. The two horizontal bands are the
      gravitational emission scale $\gammad$, and backreaction scale
      $\gammac$, for the loop distribution. Depending on $\omega$,
      they can be in or out the integration domains.}
    \label{fig:domains_freqs}
  \end{center}
\end{figure}

In figure~\ref{fig:domains_1Hz}, we have represented both
$\gammaw(\omega,z)$ and $\gammas(\omega,z,\hithstar)$ for the kink
case $\alpha=3/2$ and at frequency $f=1 \, \Hz$. The low redshift
behaviour of these functions can be obtained by Taylor expanding
Eqs.~\eqref{eq:gammaone} and \eqref{eq:gammastar} and one gets
\begin{equation}
\gammaw(\omega,z\ll1) \simeq \dfrac{4\pi}{\Ho \tnow}
\dfrac{\Ho}{\omega}\,,\qquad \gammas(\omega,z \ll 1,\hithstar) \simeq
\dfrac{1}{\Ho \tnow} \left(\hithstar z\right)^{\frac{\alpha}{\alpha-1}}
\left(\dfrac{\omega}{\Ho}\right)^{\frac{\alpha+1}{\alpha-1}}\,,
\end{equation}
where $\tnow$ is the age of the Universe. There is a crossover
redshift, $\zx$, such that for $z < \zx$ one has $\gammas < \gammaw$
and for $z>\zx$, $\gammas > \gammaw$. This redshift is solution of
\begin{equation}
  \Ho\,  \chi(\zx) (1+\zx) =
\dfrac{(4\pi)^{\frac{\alpha-1}{\alpha}}}{\hithstar}
  \left(\dfrac{\Ho}{\omega} \right)^2.
\label{eq:zx}
\end{equation}
Notice that this equation is regular for $\alpha=1$ and this allows us
to discuss the kink collision case. From the argument of the second
Heaviside function in Eq.~\eqref{eq:burstrategz}, taking the kink
collision value for $\Ctilde$ in Eq.~\eqref{eq:Ctildes}, one sees that
there is no any explicit dependency in $\gamma$ and the condition for
having only burst-like events simplifies to $z < \zx$, $\zx$ being given
by the above expression with $\alpha=1$. As a result, the kink
collision case can be viewed in figure~\ref{fig:domains_1Hz} as the
limiting situation in which the slope of the
$\gammas(\omega,z,\hithstar)$ function becomes infinite. Independently
of the loop distribution, all kink collisions close to the observer,
namely occurring at $z < \zx$ are well separated GW bursts, as opposed
to those occurring at $z>\zx$ which are accounted for as stochastic.
For cusp events, the shape of $\gammas$ is very similar to the one
plotted in figure~\ref{fig:domains_1Hz} only the slope is steeper while
$\gammaw$ remains the same for all GW source types (see figure~\ref{fig:domains_freqs}).

At large redshifts, assuming the background to be pure radiation era,
one gets
\begin{equation}
\begin{aligned}
  \gammaw(\omega,z\gg\zeq) & \simeq 8 \pi \sqrt{\OmegaR} \, z \,\dfrac{\Ho}{\omega}\,,\\
  \gammas(\omega,z\gg\zeq,\hithstar) & \simeq 2 \sqrt{\OmegaR} \left(\Ho
\chio \hithstar\right)^{\frac{\alpha}{\alpha-1}}
z^{1+\frac{\alpha}{\alpha-1}} \left(\dfrac{\omega}{\Ho}\right)^{\frac{\alpha+1}{\alpha-1}}\,.
\end{aligned}
\end{equation}
As a result, $\gammas$ grows always faster than $\gammaw$ at large
redshifts and once $z>\zx$, one always has the hierarchy
$\gammas(\omega,z,\hithstar) > \gammaw(\omega,z)$ as in
figure~\ref{fig:domains_1Hz}. Notice that the loop distribution
vanishes on super-horizon distances such that all integrals are also
vanishing for $\gamma > \gammai$. It implies that there are two other
relevant redshifts, $\zw$ and $\zs$, defined by $\gammaw(\omega,\zw) =
\gammai$ and $\gammas(\omega,\zs,\hithstar) = \gammai$, respectively.
They are the solutions of
\begin{equation}
\Ho t(\zw)(1+\zw) = \dfrac{4 \pi}{\gammai} \dfrac{\Ho}{\omega}, \qquad
\left[\dfrac{\chi(\zs)}{t(\zs)}\right]^{\alpha} \Ho t(\zs) (1+\zs) =
\dfrac{\gammai^{\alpha-1}}{\hithstar^{\alpha}}
\left(\dfrac{\Ho}{\omega} \right)^{\alpha+1}\,,
\end{equation}
and have been represented in figure~\ref{fig:domains_1Hz}. The redshift
$\zs$ is the frontier above which all GW emission events observed at
angular frequency $\omega$ will be classified as stochastic, while $\zw$ is
the maximum observable redshift at this frequency.

In figure~\ref{fig:domains_freqs}, we have represented the dependence
of the domain boundaries $\gammaw$ and $\gammas$ with respect to the
observed frequency $f=\omega/(2\pi)$ and for the cusp case
($\alpha=3$). Although the dependence of $\gammaw(\omega,z)$ is
explicit in Eq.~\eqref{eq:gammaone}, it varies as $\omega^{-1}$,
$\gammas(\omega,z,\hithstar)$ has a non-trivial behaviour due to its
dependence on the values of $\hithstar(\omega)$. For the illustration,
we have chosen values of $\hithstar(\omega)$ solving the rate equation
for a loop distribution with $\GU = 10^{-7}$. This loop distribution
has a gravitational emission scale at $\gammad \simeq 5\times10^{-6}$,
represented by a horizontal orange band in
figure~\ref{fig:domains_freqs}. The gravitational backreaction scale
$\gammac \simeq 2\times 10^{-10}$ in the matter era while $\gammac
\simeq 3 \times 10^{-9}$ in the radiation era. It is represented as a
blue discontinuous band. For most of the cosmological history, the
loop distribution is in scaling and assumes a shape as plotted in
figure~\ref{fig:ldtrans}, namely $t^4(z)\calF(\gamma,z)$ is
independent of $z$, maximal for $\gamma < \gammac$ and decreases for
$\gamma > \gammad$ to vanish at $\gamma=\gammai$. From this
observation, one can conclude without any calculation that the
stochastic gravitational wave spectrum should exhibit two
characteristic frequencies corresponding to the values of $\omega$ at
which $\gammac$ and $\gammad$ enters into the visibility domains. This
is confirmed in the next section in which we present our results.

\subsection{Results}
\label{sec:nres}

We have derived the stochastic gravitational wave spectrum
$\OmegaSGWH(\omega)$ along the lines described in the previous
section. For a given value of $\omega$, the rate equation
\eqref{eq:burstrategz} is solved for $\hithstar(\omega)$. In order to
keep track of the non-scaling loops appearing during the transition
from the radiation era to the matter era, the loop distribution $t^4
\calF$ is numerically computed at each redshift $z$ by solving the
Boltzmann equation of Ref.~\cite{Lorenz:2010sm} [see Eq.~(33) in this
  work]. The two-dimensional integrals appearing in both
Eqs.~\eqref{eq:onesgwgz} and \eqref{eq:burstrategz} have been computed
by means of the public $\texttt{CUBA}$ library~\cite{Hahn:2004fe,
  Hahn:2014fua} coupled to a Brent's method for solving the rate
equation. Thermal history effects have been included in the
cosmological variables $\chi(z)$, $H(z)$ and $t(z)$ according to the
Hindmarsh-Philipsen equation of state B~\cite{Hindmarsh:2005ix,
  Belanger:2014vza}.
\begin{figure}
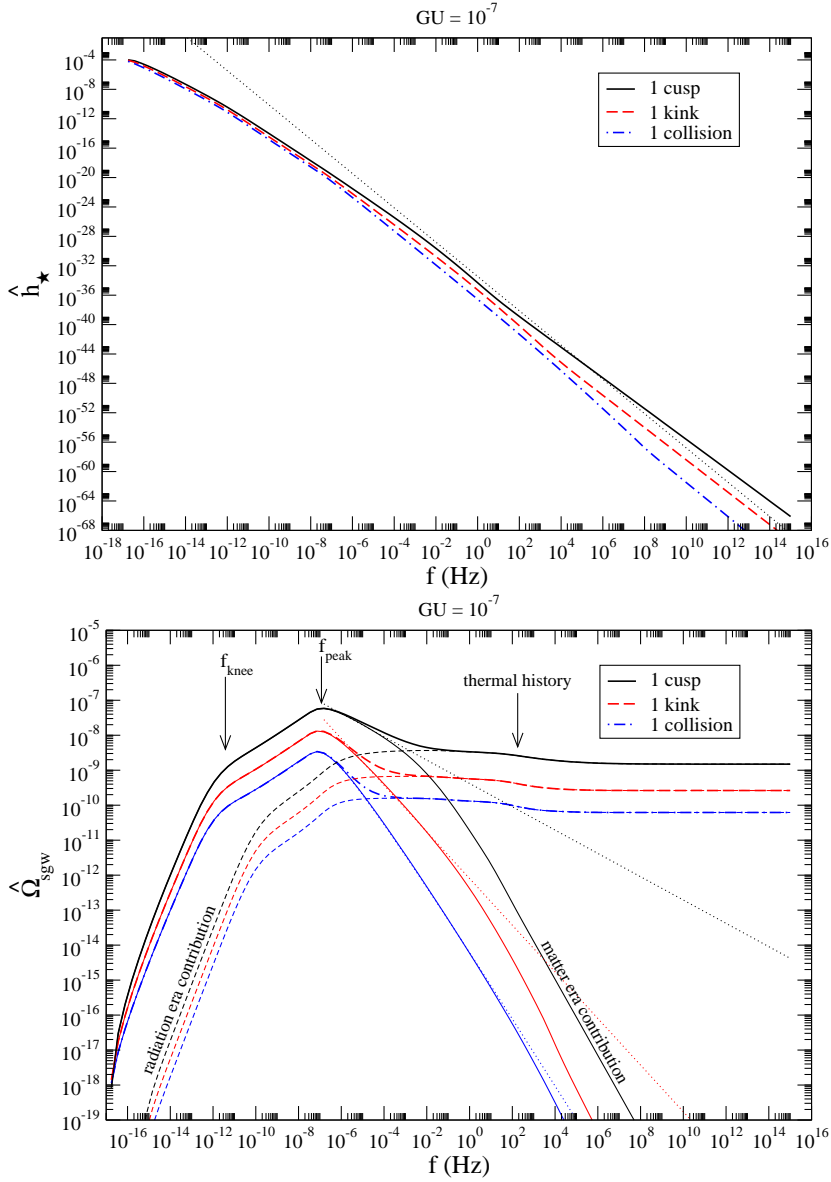

  \begin{center}
    \includegraphics[width=\onefigw]{hithstar}
    \includegraphics[width=\onefigw]{sgwhspectrum}
    \caption{The upper panel represents the threshold strain
      $\hithstar(\omega)$, solution of Eq.~\eqref{eq:burstrategz},
      under which GW events are considered stochastic as a function of
      the observed frequency $f=\omega/(2\pi)$. The thick solid line
      (black) is for one cusp, the thick dashed line (red) for one
      kink and the thick dotted-dashed line (blue) for one kink
      collision, per loop oscillation. The lower panel shows the
      stochastic gravitational wave spectrum $\OmegaSGWH(\omega)$ for
      the same three GW emission events. In addition, we have
      represented as thin solid and thin short-dashed lines the
      respective contribution of the matter era and radiation era
      loops to the overall spectrum. The peak is essentially produced
      by loops in the matter era while the high frequency plateau is
      induced by the loops in the radiation era. Dotted lines are
      analytic approximations (see text).}
    \label{fig:hatperevent}
  \end{center}
\end{figure}
The stochasticity threshold $\hithstar(\omega)$ is represented in the
upper panel of figure~\ref{fig:hatperevent} as a function of $f =
\omega/(2\pi)$ for a loop distribution having $\GU = 10^{-7}$. The
three curves correspond to GW emission by one cusp, one kink, or one
kink collision per loop oscillation. The corresponding spectra
$\OmegaSGWH(\omega)$ have been plotted in the lower panel of
figure~\ref{fig:hatperevent}. As expected, the cusp case dominates
over the kink by roughly an order of magnitude, which itself dominates
kink collision by also an order of magnitude. All these spectra are
similar in shape because they share very similar integration domains
in the plane $(\gamma,z)$. Notice however the different slopes right
to the peak. As we show below, they are a direct window on the
microscopic nature of the GW emission event. The thin solid and
short-dashed curves represent the respective contribution of GW events
coming from loops in the matter era (solid) and from loops in the
radiation era (short-dashed).

The behaviour of the stochastic strain threshold $\hithstar(\omega)$
can be understood by rewriting the left-hand side of the rate
equation~\eqref{eq:burstrategz} as
\begin{equation}
\calR = \int_0^{\zs} \ud z
\left[\dfrac{\chi(z)}{t(z)} \right]^2 \frac{\Ho}{H(z)} \dfrac{8 \pi}{(1+z)^4 [\Ho
    t(z)]^2} \int_{\max(\gammaw,\gammas)}^{\gammai} \ud \gamma
\dfrac{t^4\calF(\gamma,z)}{\gamma} \Deltab(\omega,\gamma,z).
\end{equation}
At high frequency, one expects $\gammaw(\omega,z) < \gammac$ such that
the rate $\calR$ is dominated by small loops at small redshift (see
figure~\ref{fig:domains_freqs}). With such a crude approximation,
replacing the beam with Eq.~\eqref{eq:beam}, one gets
\begin{equation}
\calR \propto (4\pi)^{3-\frac{3}{\alpha}}
\, \dfrac{\tfourFo}{\hithstar^3} \left(\dfrac{\Ho}{\omega} \right)^{6},
\label{eq:rateapprox}
\end{equation}
where $\order{1}$ factors have been dropped and with $\tfourFo \equiv
\tnow^4 \calF(\gamma < \gammac,z=0)$. Solving $\calR = \omega/(2\pi
\Ho)$ for $\hithstar$ gives
\begin{equation}
\hithstar(\omega) \propto (4 \pi)^{1-\frac{1}{\alpha}} \left(\tfourFo\right)^{1/3}
  \left(\dfrac{\Ho}{\omega} \right)^{7/3}.
\label{eq:hithstarapprox}
\end{equation}
This approximate expression for $\hithstar$ has been represented as
dotted lines in the upper panel of figure~\ref{fig:hatperevent} and
roughly reproduces the high frequency behaviour of $\hithstar(\omega)$
but overestimates the correct value at low frequencies.

For all sources, cusp, kink and collision, the spectrum
$\OmegaSGW(\omega)$ exhibits a knee at angular frequency $\omegaknee$ followed
by a maximum at $\omegapeak$. As mentioned in
section~\ref{sec:domains}, these two features are the result of the
scales $\gammad$ and $\gammac$ entering the loop visibility
domains. From figure~\ref{fig:domains_freqs}, a sufficient condition
for $\gammac$ to belong to the integration domain is
$\gammaw(\omega,z=0) \le \gammac$. Defining the angular frequency $\omegac$ at
which these two quantities are equal, one gets
\begin{equation}
\dfrac{\omegac}{\Ho} = \dfrac{4\pi}{\tnow \Ho \gammac}\,.
\label{eq:omegac}
\end{equation}
For $\GU=10^{-7}$ and $\Upsilon=20$, one has $\omegac/(2\pi) \simeq 2
\times 10^{-8}\,\Hz$ which gives the correct order of magnitude for
the peak frequency, see figure~\ref{fig:hatperevent}. However, a
closer look to this figure reveals that the peak frequency depends on
the type of GW burst, namely on the value of $\alpha$. In order to
derive a better analytic approximation, let us consider
$\OmegaSGWH(\omega)$ around the maximum. From Eqs~\eqref{eq:onesgwgz}
and \eqref{eq:hatsdef}, it reads
\begin{equation}
\OmegaSGWH(\omega) = \dfrac{2 (\GU)^2 \omega^3}{3 \Ho^3}
\int_{\zx}^{\zw} \dfrac{\ud z}{(1+z)^2} \dfrac{\Ho}{H(z)}
\int_{\gammaw}^{\min(\gammas,\gammai)} \dfrac{\ud \gamma}{\gamma} t^4
  \calF(\gamma,z)\Chat^2(\omega,\gamma,z)\Deltab(\omega,\gamma,z).
\label{eq:omegasgwh}
\end{equation}
As before, we can make the crude assumption that the integral over
$\gamma$ is dominated by small loops at small redshift, and we keep
only terms having $\gamma \le \gammac$ and $z < \zcc$, where $\zcc$ is
an arbitrary cut-off. In order to keep the redshift terms of the
integral simple, let us chose $\zcc$ to be the redshift under which
the cosmological constant dominates, i.e. $H(z<\zcc) \simeq \Ho
\sqrt{\OmegaL}$. One gets
\begin{equation}
\begin{aligned}
  \OmegaSGWH(\omega) & \propto (\GU)^2 \left(\dfrac{\Ho}{\omega}
\right)^{\frac{1}{\alpha}} \zcc \dfrac{(4 \pi)^{1-\frac{1}{\alpha}}}{(\tnow
  \Ho)^{3+\frac{1}{\alpha}}} \,  \tfourFo \,
\dfrac{\gammac^{1-\frac{1}{\alpha}}}{1-\frac{1}{\alpha}} \left\{1 -
  \left[\dfrac{\gammaw(\omega,0)}{\gammac}
    \right]^{1-\frac{1}{\alpha}} \right\} \\
& \simeq (\GU)^2 \dfrac{\tfourFo}{(\tnow \Ho)^{3+\frac{1}{\alpha}}}
  \dfrac{(4\pi\gammac)^{1-\frac{1}{\alpha}}}{1-\frac{1}{\alpha}}
  \left(\dfrac{\Ho}{\omegac}\right)^{\frac{1}{\alpha}}
  \left[\left(\dfrac{\omegac}{\omega}\right)^{\frac{1}{\alpha}} -
    \dfrac{\omegac}{\omega} \right],
\label{eq:omegasgwhlowz}
\end{aligned}
\end{equation}
in which we have dropped all $\order{1}$ factors and used the
definition~\eqref{eq:omegac} of $\omegac$. This formula is valid only
for $\omega > \omegac$. In spite of the crude approximations made, the
above expression is surprisingly accurate. The dependency in $\omega$
describes a maximum occurring at
\begin{equation}
\dfrac{\omegapeak}{\Ho} \simeq \dfrac{4 \pi
  \alpha^{\frac{\alpha}{\alpha-1}}}{\tnow \Ho \gammac} =
\dfrac{4\pi \alpha^{\frac{\alpha}{\alpha-1}}}{\tnow \Ho \Upsilon} \left(\GU\right)^{-(1+2\chi)} \,.
\label{eq:omegapeak}
\end{equation}
Notice that all the above expressions are regular for $\alpha
\rightarrow 1$ and can be used for kink collisions as well: one has
$\alpha^{\alpha/(\alpha-1)} \rightarrow e$ while the power laws in
Eq.~\eqref{eq:omegasgwhlowz} becomes logarithmic
functions. Equation~\eqref{eq:omegapeak} for the peak frequency can be
compared to the exact numerical results of
figure~\ref{fig:hatperevent}. One finds $\omegapeak/(2\pi) = 1.3\times
10^{-7}\,\Hz$ which matches the exact numerical value. For kinks,
$\omegapeak/(2\pi) = 8.3 \times 10^{-8}\,\Hz$ for an exact value at
$8.7 \times 10^{-8}$ and for collisions one gets $\omegapeak/(2\pi) =
6.7 \times 10^{-8}\,\Hz$ instead of $7.5 \times 10^{-8}\,\Hz$.
The $\omega$-dependency of Eq.~\eqref{eq:omegasgwhlowz} around the
maximum is such that
\begin{equation}
\left. \OmegaSGWH(\omega \gtrsim \omegapeak)\right|_{\alpha\ne 1}
\propto \omega^{-\frac{1}{\alpha}}, \qquad \left. \OmegaSGWH(\omega \gtrsim
\omegapeak)\right|_{\alpha = 1} \propto \omega^{-1} \ln \omega.
\end{equation}
These analytic approximations have been represented in the lower
panel of figure~\ref{fig:hatperevent} as dotted curves for $\omega >
\omegac$ and appear to fit relatively well the decrease of
$\OmegaSGWH(\omega)$ just after its maximum. As a result, there is a
small window in frequencies, for $\omega \gtrsim \omegapeak$, in which
the slope of the stochastic gravitational wave spectrum directly
reflects the type of GW bursts occurring along the cosmic string
loops. Plugging Eq.~\eqref{eq:omegapeak} into Eq.~\eqref{eq:omegasgwhlowz}
allows us to derive the dependency of the maximal power with respect
to the string tension. One gets
\begin{equation}
\OmegaSGWH(\omegapeak) \propto (\GU)^2 \, \tfourFo \, \gammac \propto
(\GU)^2 \, \dfrac{\gammac^{2\chi-1}}{\gammad} \propto \left(\GU\right)^{4 \chi^2},
\label{eq:omegasgwhpeak}
\end{equation}
where we have used the approximate relation $\tfourFo \propto
\gammac^{2\chi-2}/\gammad$ derived in
Ref.~\cite{Lorenz:2010sm}. Since, for not too small values of $\GU$,
the peak is dominated by matter era loops, the value $\chi=\chimat$
should be used in this expression.

Concerning the knee in the spectrum, it corresponds to the change of
behaviour of the loop distribution at $\gamma=\gammad$ (see
figure~\ref{fig:ldtrans}). As a result, all of the previous
approximations can equally be applied to determine the knee frequency
by formally replacing $\gammac$ by $\gammad$. One obtains
\begin{equation}
\omegaknee \simeq \dfrac{4 \pi
  \alpha^{\frac{\alpha}{\alpha-1}}}{\tnow\Ho \gammad} =  \dfrac{4 \pi
  \alpha^{\frac{\alpha}{\alpha-1}}}{\tnow\Ho \Gamma} (\GU)^{-1}\,,
\label{eq:omegaknee}
\end{equation}
with
\begin{equation}
\OmegaSGWH(\omegaknee) \propto (\GU)^2 \, \tnow^4 \calF(\gammad,0) \,
\gammad \propto (\GU)^2 \gammad^{2\chi-2} \propto (\GU)^{2 \chi}\,.
\end{equation}
These approximations cease however to be valid as soon as the
radiation-era loops contribute significantly to the peak.

The radiation-era loops are expected to be the main source of GW in
the high frequency limit. Figure~\ref{fig:domains_freqs} indeed shows
that, at high frequency, the domains in which the loop distribution is
maximal, namely for $\gamma \le \gammac$, end up being in the
radiation-era. This can be explicitly seen by plugging
Eq.~\eqref{eq:hithstarapprox} into the definition of $\zx$ in
Eq.~\eqref{eq:zx}:
\begin{equation}
  \zx \propto
  \left(\tfourFo\right)^{-\frac{1}{3}} \left(\dfrac{\omega}{\Ho} \right)^{\frac{1}{3\alpha}}.
\end{equation}
For $\omega/\Ho$ large enough, $\zx$ lies in the
radiation-era. Keeping only loops having $\gamma \le \gammac$ and
taking all the redshift dependent terms in Eq.~\eqref{eq:omegasgwh} to
be in the radiation era, one recovers that the high frequency limit is
frequency independent~\cite{Olmez:2010bi}
\begin{equation}
\OmegaSGWH(\omega) \propto \OmegaR \rdof(\zwc)
\frac{\gammac^{2\chi-1}}{\gammad} \propto \OmegaR \rdof(\zwc) \, \left(\GU\right)^{4\chi^2},
\label{eq:omegasgwhinfty}
\end{equation}
where $\zwc$ is a redshift at which $\gammaw(\zwc)=\gammac$ is
satisfied.  The quantity $\rdof(z) \equiv \gszero^{4/3}
\g(z)/[\gs(z)^{4/3} \gzero]$ denotes the change in energetic ($\g$)
and entropic ($\gs$) relativistic degrees of freedom between $z$ and
today. The presence of $\rdof(\zwc)$ justifies the smooth steps
visible in figure~\ref{fig:hatperevent} and labelled as ``thermal
history''. Depending on the redshift $z$ at which the GW event has
been emitted, the amplitude of the received signal is modulated by
$\rdof(z)$, through its effect on $H(z)$, $\chi(z)$ and $t(z)$. As a
result, and as it is the case for GW produced during inflation, cosmic
string loops could be used to perform GW tomography of the
radiation-era~\cite{Watanabe:2006qe, Kuroyanagi:2013ns}. Although the
amplitude of the high frequency plateau scales as $(\GU)^{4 \chi^2}$,
which is the same function as the peak amplitude in
Eq.~\eqref{eq:omegasgwhpeak}, one has to use the radiation era value
$\chi=\chirad$ in Eq.~\eqref{eq:omegasgwhinfty}. Therefore, the peak
and the plateau do not scale in the exact same way with respect to
$\GU$.

\begin{figure}
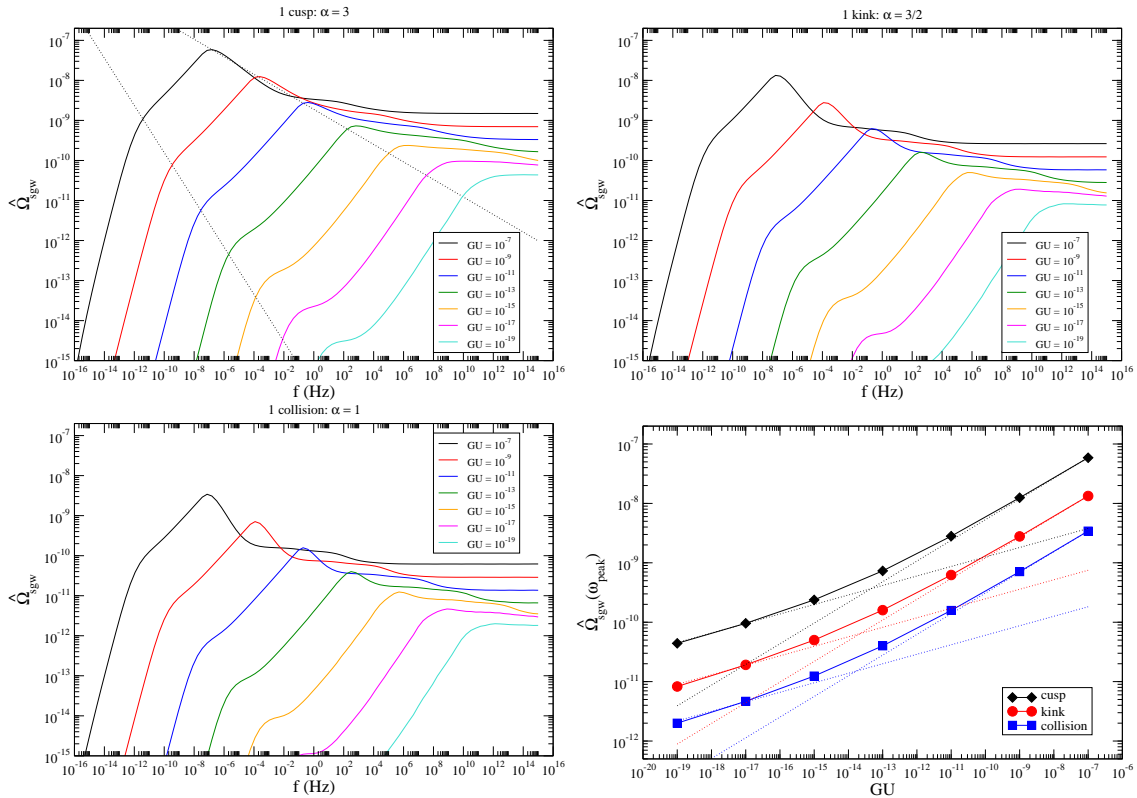

  \begin{center}
    \includegraphics[width=\twofigw]{sgwhmax_cusp}
    \includegraphics[width=\twofigw]{sgwhmax_kink}
    \includegraphics[width=\twofigw]{sgwhmax_collision}
    \includegraphics[width=\twofigw]{sgwhmax_gu}
    \caption{Stochastic GW spectra for one cusp (upper left), one kink
      (upper right) and one collision (lower left) event per loop
      oscillation as a function of the string tension $\GU$. The
      dotted lines in the upper left panel are the analytic
      approximations of Eqs.~\eqref{eq:omegapeak} and
      \eqref{eq:omegasgwhpeak}. The lower right panel shows the
      maximal amplitude $\OmegaSGWH(\omegapeak)$ as a function of
      $\GU$ together with the analytic approximations as dotted lines
      (see text).}
    \label{fig:GUdep}
  \end{center}
\end{figure}

In figure~\ref{fig:GUdep}, we have plotted various spectra
$\OmegaSGWH(\omega)$, still for one cusp, one kink and one collision
event per loop oscillation, for different values of the string tension
$\GU$. For the cusp case (upper right panel) the dotted lines are the
analytic approximations defined in Eqs.~\eqref{eq:omegapeak} and
\eqref{eq:omegasgwhpeak}. As can be seen on this plot, the
approximations are accurate down to $\GU \simeq 10^{-11}$ while for
lower string tension we find that Eq.~\eqref{eq:omegasgwhpeak}
underestimates the peak amplitude. Let us stress than
Eq.~\eqref{eq:omegapeak} for the peak frequency remains accurate. In
the lower right panel of figure~\ref{fig:GUdep}, we have plotted the
numerical value of $\OmegaSGWH(\omegapeak)$ as a function of $\GU$. In
this plot, the dotted lines are the approximations of
Eq.~\eqref{eq:omegasgwhpeak} using either $\chi=\chimat$ (steeper
slope) or $\chi=\chirad$. Very low string tensions are better fitted
with the radiation era value for $\chi$ suggesting that radiation-era
loops significantly contribute to the overall peak. The shape of the
spectrum for these values of $\GU$ is also severely distorted: the
maximum occurs at very high frequency and it progressively becomes
undistinguishable from the plateau. Notice however the different
behaviour of the knee frequency and amplitude.

\subsection{Comparison with previous works}
\label{sec:pw}

Because we have paid special attention to consider a realistic loop
distribution, it is difficult to compare our results to those having
used single-sized distributions or production
functions~\cite{Caldwell:1991jj, Damour:2000wa, DePies:2007bm,
  Regimbau:2011bm, Binetruy:2012ze, Kuroyanagi:2012wm, Aasi:2013vna,
  Henrot-Versille:2014jua, Sousa:2016ggw}. However, our results could
be compared to those of Refs.~\cite{Olmez:2010bi, Sanidas:2012ee,
  Blanco-Pillado:2013qja} which have considered non-trivial loop
distributions. The overall shape of our spectrum matches with the one
presented in these works, namely all of them exhibit a plateau at high
frequencies, a maximum and a fast decay at low frequencies. Concerning
the relative amplitude of cusp versus kink, our spectra have a more
pronounced hierarchy that the ones computed in
Ref.~\cite{Olmez:2010bi} but this could be the result of some specific
numerical values chosen for their coefficient $c_\alpha$ in
Eq.~\eqref{eq:calpha}. The presence of a knee in the spectrum is
somehow reminiscent with the two-scales loop model of
Ref.~\cite{Sanidas:2012ee}, although our spectra are significantly
different in shapes and amplitude in this case. Because
Ref.~\cite{Sanidas:2012ee} uses more than five parameters to model the
spectrum, it is possible that some combination of them, and a
multi-scale loop model, could reproduce our result but we have not
attempted further comparison.

\begin{figure}
  \begin{center}
    \includegraphics[width=\onefigw]{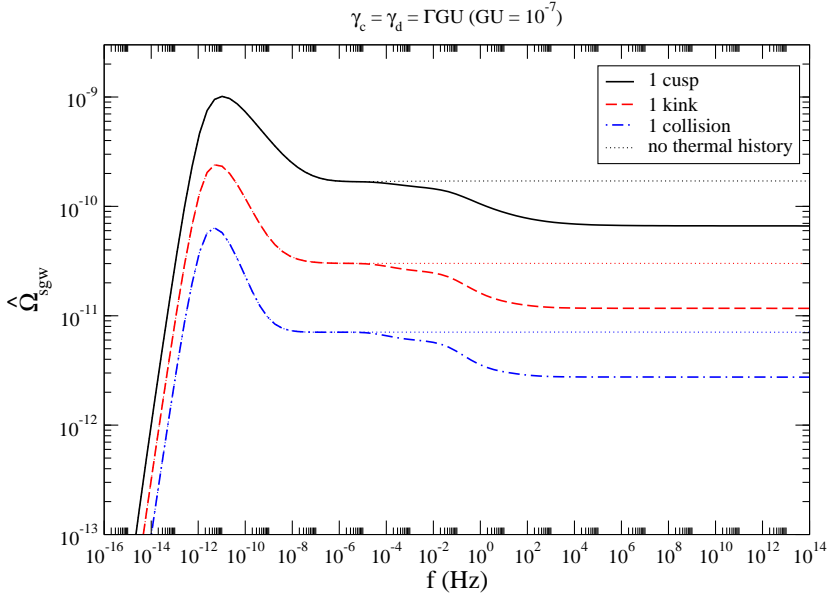}
    \caption{Stochastic GW spectra for one cusp, one kink and one
      collision event per loop oscillation for a scaling loop
      distribution having $\gammac = \gammad = \Gamma \GU$ with
      $\Gamma=50$ and $\GU=10^{-7}$. Notice that the physical spectra
      are $\OmegaSGW(\omega) = c_\alpha^2 \OmegaSGWH(\omega)$ where
      $c_\alpha$ is given in Eq.~\eqref{eq:calpha}. The dotted curves
      are derived \emph{without} thermal history by assuming a pure
      radiation era without any changes in the number of relativistic
      degrees of freedom.}
    \label{fig:gammacisd}
  \end{center}
\end{figure}

More interestingly the results of Ref.~\cite{Blanco-Pillado:2013qja}
are based on a realistic loop distribution inferred from Nambu-Goto
simulations. Their spectrum has been derived without thermal history
effects and for cusp events only. Up to the thermal effects in the
high frequency part, the main differences are that the peak of their
spectrum matches the knee in ours. As discussed earlier, the overall
maximum of $\OmegaSGWH(\omega)$ is associated with the gravitational
backreaction scale $\gammac$ in the loop distribution whereas the knee
traces the gravitational emission length scale $\gammad$. The results
of Ref.~\cite{Blanco-Pillado:2013qja} have been derived under the
assumption $\gammac=\gammad=\Gamma \GU$. In
figure~\ref{fig:gammacisd}, we have therefore plotted the cusp, kink
and collision dimensionless spectra $\OmegaSGWH(\omega)$ in the
particular case $\gammac=\gammad$ and without thermal history
effects. As expected, the peak and knee previously visible in
figure~\ref{fig:hatperevent} are merged and the cusp spectrum can be
directly compared to the one derived in
Ref.~\cite{Blanco-Pillado:2013qja} (see figure~13) provided we
estimate $c_{3}$. From Eq.~\eqref{eq:calpha}, taking $\beta=1$ and
$\etas$ from Eq.~\eqref{eq:etas}, one gets $c_{3}\simeq 7$ such that
one has to multiply the cusp spectrum of figure~\ref{fig:gammacisd} by
$c_3^2$ to get the physical one. For $\GU=10^{-7}$, we get the peak
frequency at $\omegapeak/(2\pi) \simeq 1.0 \times 10^{-11}\,\Hz$,
which, up to a factor of two, matches the one of figure~13 in
Ref.~\cite{Blanco-Pillado:2013qja}. For the maximum amplitude, we get
$\OmegaSGW(\omegapeak) \simeq 5\times 10^{-8}$, to be compared to a
value around $6\times 10^{-8}$ as it can be inferred from their
figure~13. Both spectra are therefore within a ten percent agreement
for the amplitude when derived under the condition
$\gammad=\gammac$. Let us notice however that the power law exponents
of the loop distribution between Ref.~\cite{Ringeval:2005kr} and
Ref.~\cite{Blanco-Pillado:2013qja} are slightly different. In terms of
our parameter $\chi$, for the radiation era,
Ref.~\cite{Blanco-Pillado:2013qja} uses a value of $\chirad=0.25$
(and $\chimat=0.5$) such that this agreement is degraded for the lower
values of $\GU$. For instance, for $\GU=10^{-11}$, we find an
agreement only up to a factor of $6$ for the amplitude. For much lower
values of $GU$, as can be checked in our figure~\ref{fig:GUdep},
relaxation effects, which are encoded in our Boltzmann approach,
becomes significant and the loop distributions cannot be
straighforwardly compared anymore.

Although this agreement can be considered as a validation of our
numerical method, let us stress the importance of considering the two
gravitational length scales $\gammad$ and $\gammac$ separated. As can
be seen in figure~\ref{fig:GUdep}, the overall peak of our spectrum is
significantly higher than the knee while its amplitude decreases with
$\GU$ much slower than the amplitude at the knee frequency. This is
expected from the shape of the loop distribution in
figure~\ref{fig:ldtrans}. The size and number of loops of size
$\gammac$ is significantly boosted compared to those of size $\gammad$
when $\GU \rightarrow 0$. Let us also stress the importance of thermal
history effects which lower the plateau by a factor $\order{3}$,
especially around the LIGO frequencies.

Finally, up to our knowledge, the collision spectrum has not been
derived before. Its importance will be discussed in the next section
in which we explore the range of remaining theoretical uncertainties
on the loop microstructure.

\section{Loop microstructure and constraints}
\label{sec:micro}

The string microstructure affects the observable GW spectrum through
the coefficients $c_\alpha$ in Eq.~\eqref{eq:calpha}. The precise
determination of these coefficients is out of the scope of the current
work and would require a precise examination of various correlation
functions along the string worldsheet. However, we expect the order
of magnitude for $c_\alpha$ to be relatively accurate and in the
following we will be using the fiducial values
\begin{equation}
  \beta=1, \qquad v_\pm^2 = 0.5,
\end{equation}
and $\etas$ given by Eq.~\eqref{eq:etas}. As discussed in the
introduction, the number of cusps, kinks and collisions per loop
oscillations is currently unknown and constitutes the main source of
theoretical uncertainties on the observable spectrum. Denoting by
$\Ncusp$, $\Nkink$ and $\Ncoll$ the number of cusps, kinks and
collisions appearing per loop oscillation, the final spectrum
$\OmegaSGWH(\omega)$ can be obtained by the weighted contributions of
each of the spectra derived in the previous section. Notice however
that the stochasticity threshold $\hithstar$ has also to be recomputed
for each value of $\Ncusp$, $\Nkink$ and $\Ncoll$ and that these
numbers may depend on the loop size $\gamma$.

\subsection{Gravitational wave emission bound}
\label{sec:Nbounds}

The numbers $\Ncusp$, $\Nkink$ and $\Ncoll$ cannot take arbitrarily
large values as the total power emitted by each loop under the form of
GW is given by $\Pgw = \Gamma \GUU$. For one GW emission event per
loop oscillation, in the local wave zone around each loop, one has
$\Pgw =\rhogw^{(\circ)} 4\pi r^2 \Deltab$. From Eq.~\eqref{eq:rhogws}, one gets
\begin{equation}
\Pgw= \dfrac{\GUU}{\ell^2} \sum_{n=1}^{+\infty} \varpi_n^2 \,
\Cbar^2(\varpi_n,\ell) \Deltab(\varpi_n,\ell).
\end{equation}
This expression is valid for any sources, but one expects the cusp,
kink or collision waveforms to dominate over all the other terms in
the high frequency regime, i.e., for $n \gg 1$, see
section~\ref{sec:ckandkk}. On the contrary, for the first values of
$n$, one expects $\Cbar^2$ to be dominated by the loop oscillation
modes. Therefore, we define a threshold, $\ncut$, above which GW
emission is dominated by cusp, kink or collision events only. A
precise value for $\ncut$ is not known, although a possible number
inferred from the simulations of Ref.~\cite{PhysRevD.45.1898} could be
taken as $\order{10}$. In the following, we will be using the safe value
$\ncut=5$, which might underestimate the maximal allowed numbers of cusp,
kinks and collision events per loop oscillation. As a result, the
GW power emitted by the string microstructure only (cusp, kinks or collisions) per loop
oscillation reads
\begin{equation}
\Pgw^{(\alpha)} = N_{(\alpha)} \dfrac{\GUU}{\ell^2} \sum_{n=\ncut}^{+\infty}
\Cbar^2_{(\alpha)}(\varpi_n,\ell) \Deltab^{(\alpha)}(\varpi_n,\ell),
\end{equation}
where $N_{(\alpha)}$, $\Cbar_{(\alpha)}^2$ and $\Deltab^{(\alpha)}$ are for the cusp, kink and
collision waveforms ($N_{(3)}\equiv\Ncusp$, $N_{(3/2)}\equiv\Nkink$,
$N_{(1)}\equiv\Ncoll$). Because $\Pgw^{(\alpha)} \le \Pgw$, using
Eqs.~\eqref{eq:beam} and \eqref{eq:Ctildes}, one obtains
\begin{equation}
N_{(\alpha)} \le \left(\dfrac{\sqrt{3}}{2}\right)^{1-\frac{1}{\alpha}}
\dfrac{(4\pi)^{\frac{2}{\alpha}} \, \Gamma}{b_\alpha \, c_\alpha^2
\displaystyle \sum_{n=\ncut}^{+\infty} \dfrac{1}{n^{1+\frac{1}{\alpha}}}}\,.
\end{equation}
Using the fiducial values mentioned before, with $\Gamma=50$, one
gets
\begin{equation}
\Ncusp \le 11, \qquad \Nkink \le 257, \qquad \Ncoll \le 4459.
\end{equation}
As can be seen in figure~\ref{fig:hatperevent}, a large number of
kinks or collision events per loop oscillation may therefore allow the
total stochastic GW background to be dominated by kinks or collisions
instead of the naively expected cusps. For this reason, in the
following, we devise four prototypical scenarios according to the
hierarchy between the number of cusps and kinks present on the loops.

\subsection{Prototypical models of string microstructure}
\label{sec:proto}

\begin{figure}
  \begin{center}
    \includegraphics[width=\onefigw]{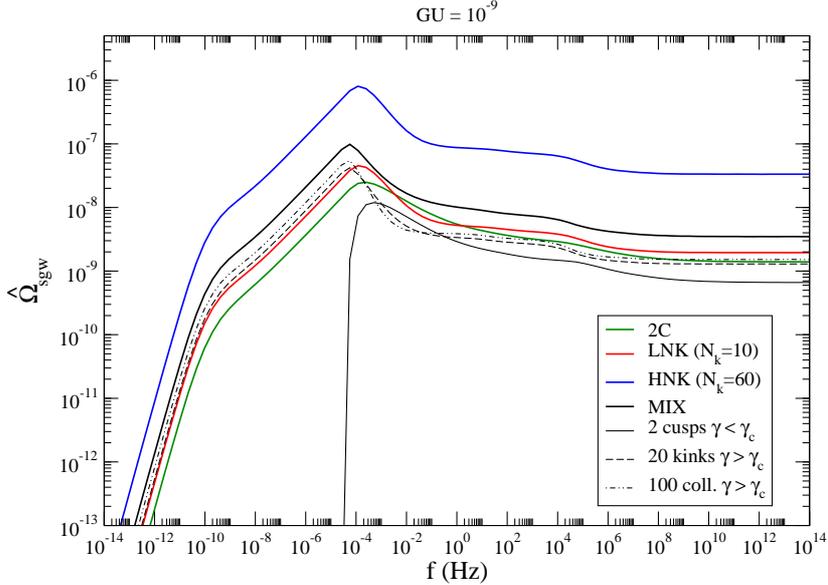}
    \caption{Stochastic GW spectra for the prototypical models having
      two cusps (2C), a low number of kinks $\Nkink=10$ (LNK), and a
      high number of kinks $\Nkink = 60$ (HNK) per loop
      oscillation. The scenario labelled as MIX has two cusps for loops
      smaller than the gravitational backreaction length $\gamma \le
      \gammac$ and only kinks for larger loops (here $\Nkink=20$).}
    \label{fig:proto}
  \end{center}
\end{figure}

In the following, we will always assume that kinks are created in
pairs, as it is observed in Nambu-Goto simulations such that $\Ncoll =
\Nkink^2/4$. Under this assumption, the number of kinks per loop
oscillation is actually bounded by the maximal numbers of collisions
and one gets $\Nkink \le 133$.

The first prototypical model assumes $\Ncusp=2$ and $\Nkink=0$. It is
motivated by the results of Ref.~\cite{Blanco-Pillado:2015ana}. If GW
backreaction is a very efficient process, loops microstructure is
erased and the smooth loops tend to develop two cusps per
oscillations. If the larger loops do not have too many kinks, $\Nkink
\ll \order{10}$, the overall stochastic GW background is dominated by
cusps only. One could consider this case as the standard lore. We will
refer to this model as 2C (2 cusps).

The second scenario assumes $\Ncusp=0$ and a not too large number of
kinks: $\Nkink^2/4 \ll \order{10^{-1}} \Nkink$. This condition ensures
that the stochastic GW background is always dominated by kinks and not
by collisions. We choose $\Nkink < 20$. Such a scenario may be
motivated by the results of Refs.~\cite{Wachter:2016rwc,
  Wachter:2016hgi} and would describe the cases for which GW
backreaction does not smooth the string microstructure. We will refer
to this model as ``LNK'' (low number of kinks). The third model is the
same as the previous one but assumes $20 \le \Nkink \le 133$ such that
the stochastic GW spectrum is dominated by collision events. This
model will be refereed to as ``HNK'' (high number of kinks).

Finally, we also discuss a model in which loops smaller than the
gravitational backreaction length scale $\gamma \le \gammac$ are
smooth and develop two cusps per oscillation whereas the larger ones
$\gamma > \gammac$ have only kinks. For those, we simply assume that
$\Nkink \le 133$ and this scenario may be viewed as a scale-dependent
mixture of 2C, LNK and HNK. It will be referred to as ``MIX'' in the
following.

In figure~\ref{fig:proto}, we have plotted the typical stochastic GW
spectra $\OmegaSGWH$ for these four models. Their shape can be
understood from the spectra of individual events derived in
section~\ref{sec:nres}. For the MIX scenario, we have separated the
contribution coming from cusps, kinks and collisions. Because the low
frequency part of the spectrum is generated by large loops, the cusp
contribution rapidly vanishes for $\omega < \omegapeak$.

Because $\Nkink$ can vary over two orders of magnitude between these
models, it is not possible to extract an unique model independent
constraint on $\GU$ from the current stochastic GW limits. For this
reason, in the next section, we perform a Bayesian analysis of the
parameter space $(\GU,\Nkink)$ for each of these
scenarios. Marginalising over $\Nkink$ within the appropriate prior
range allows us to obtain a robust constraint on $\GU$.

\subsection{Observational constraints}
\label{sec:mcmc}

\begin{figure}
  \begin{center}
    \includegraphics[width=\twofigw]{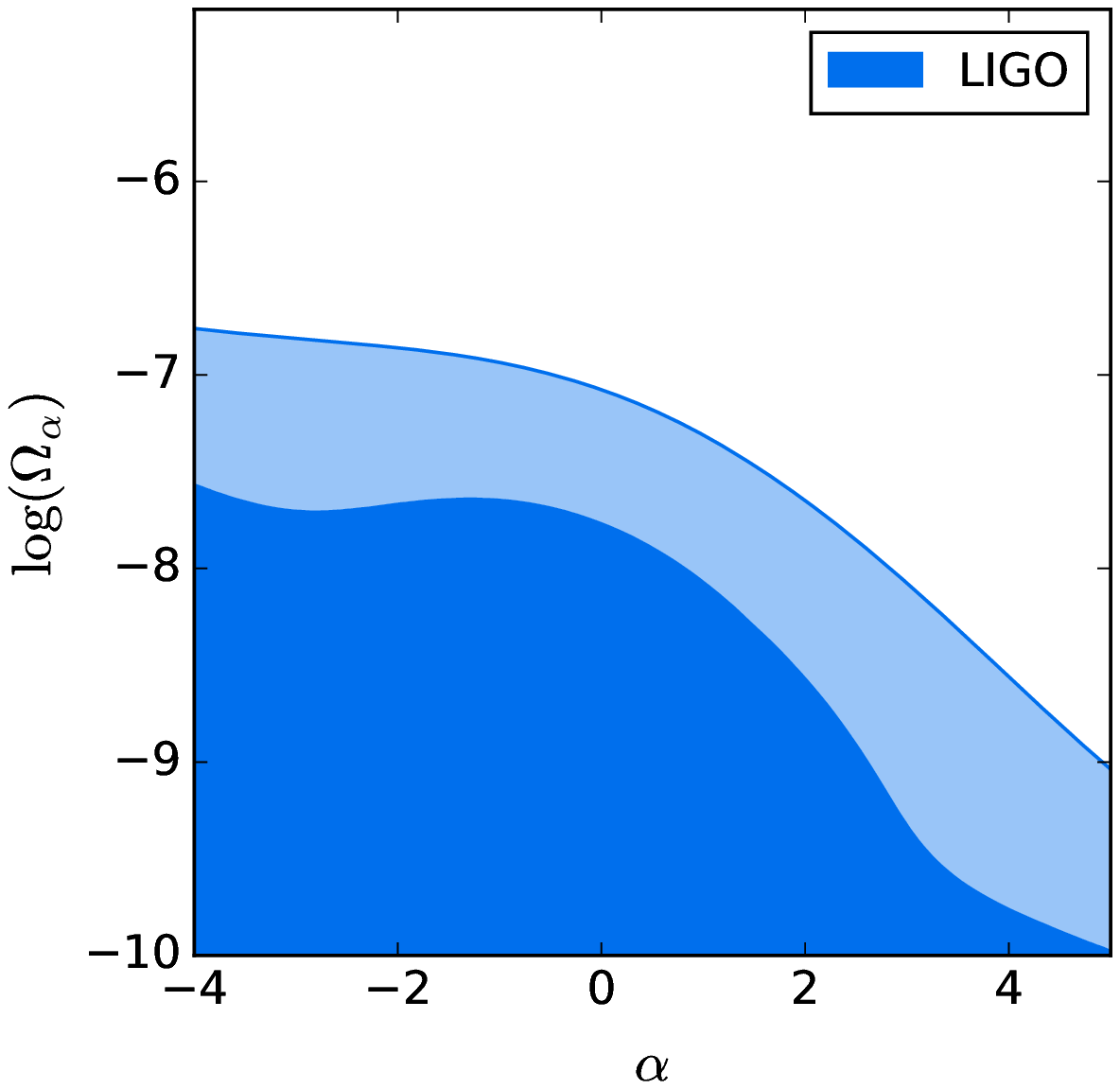}
    \includegraphics[width=\twofigw]{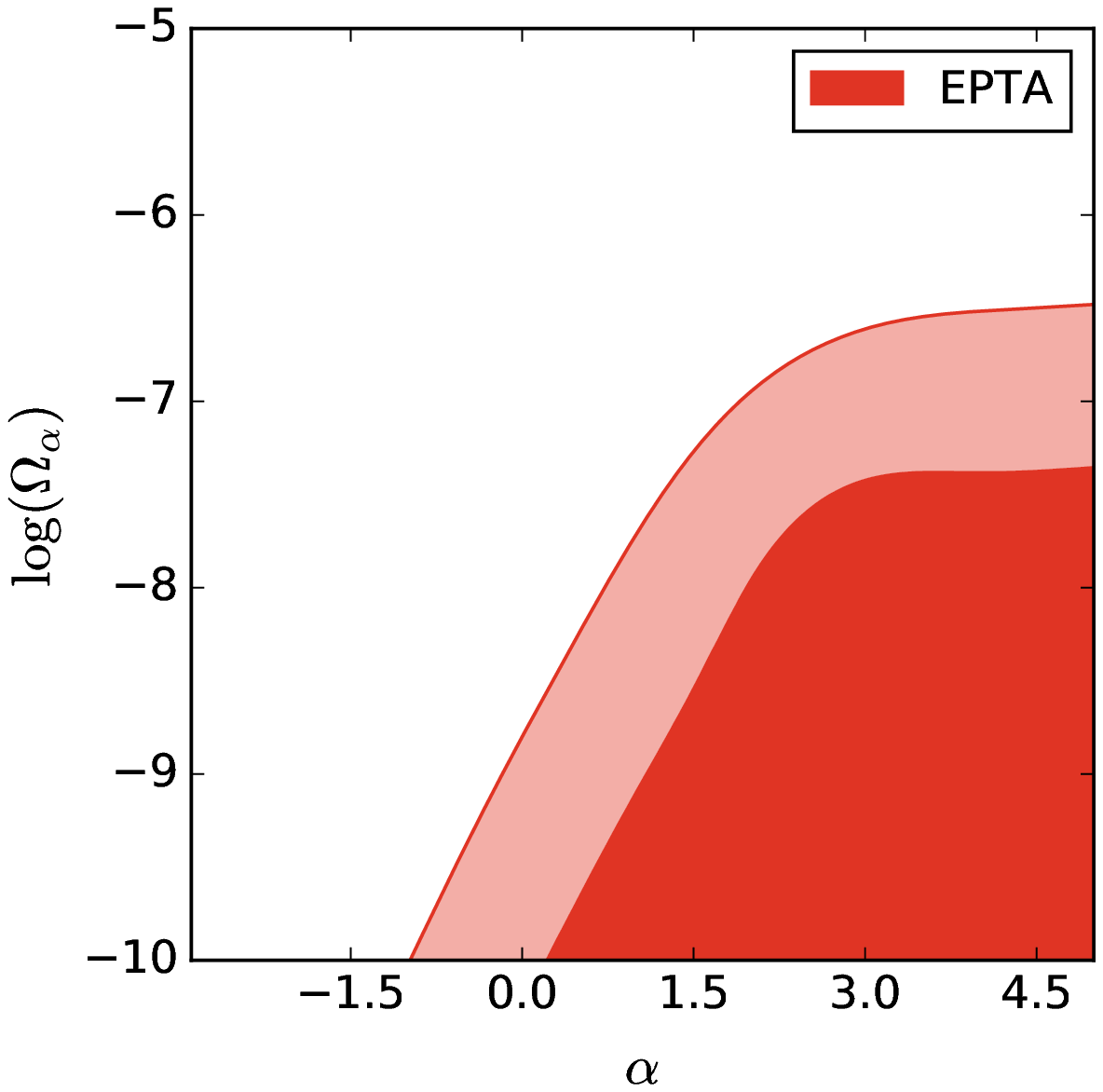}
    \caption{One- and two-sigma contour of the two-dimensional
      posterior distribution obtained by nested sampling of our toy
      LIGO (left) and EPTA (right) likelihoods. The resulting
      confidence intervals are found to reproduce well the results of
      Refs.~\cite{TheLIGOScientific:2016dpb,Verbiest:2016vem}.}
    \label{fig:ligoepta_2D}
  \end{center}
\end{figure}

The prototypical models of string microstructure presented in the
previous section have been compared to the existing bounds on the
stochastic gravitational wave background. For this purpose, we have
used the recent results of the LIGO collaboration at $f=25\,\Hz$
presented in Ref.~\cite{TheLIGOScientific:2016dpb}, as well as the
ones coming from the European Pulsar Timing Array (EPTA) at
$f=31.7\,\nHz$ in Ref.~\cite{Verbiest:2016vem}.

\begin{figure}
  \begin{center}
    \includegraphics[width=\twofigw]{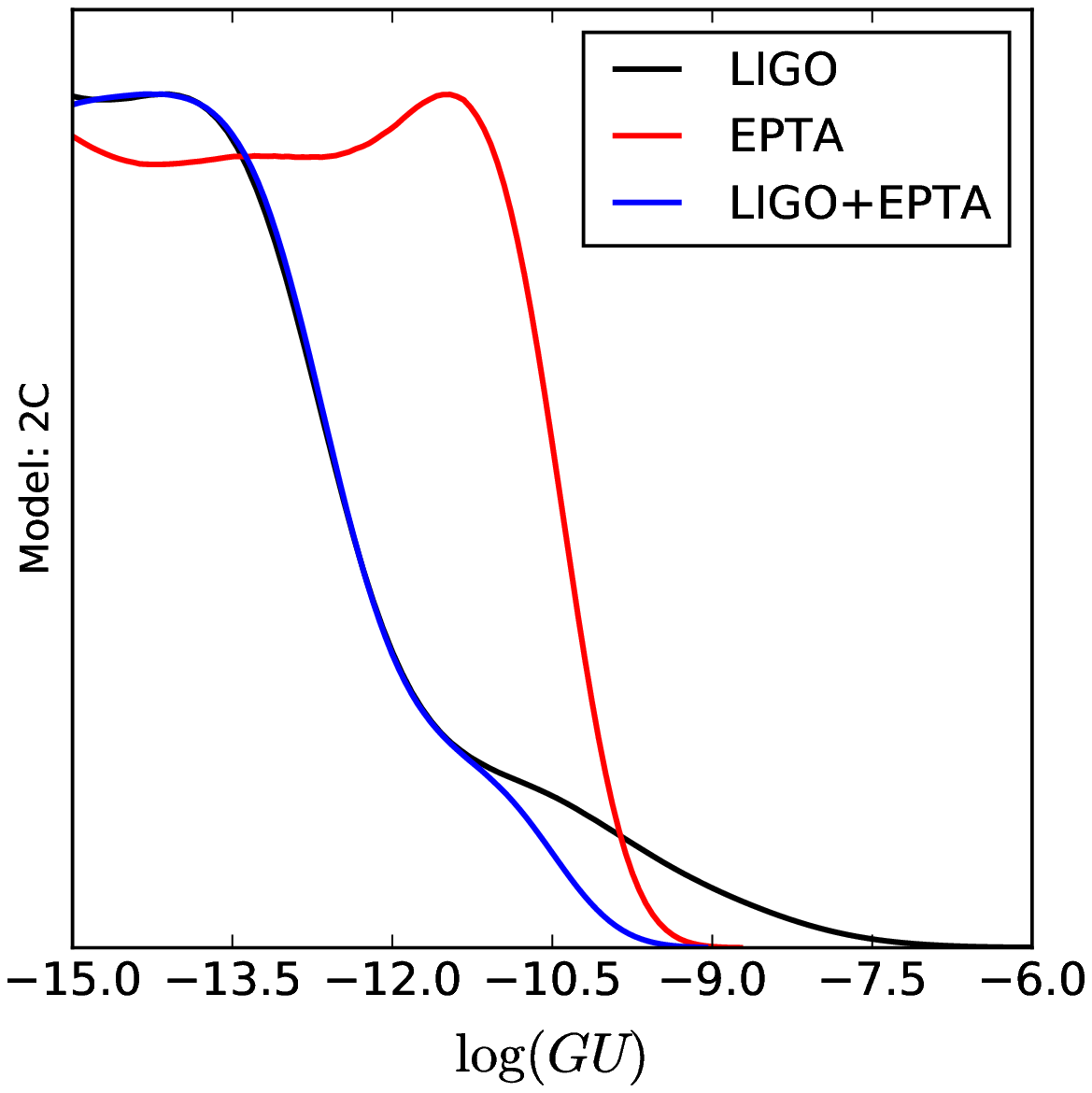}
    \includegraphics[width=\twofigw]{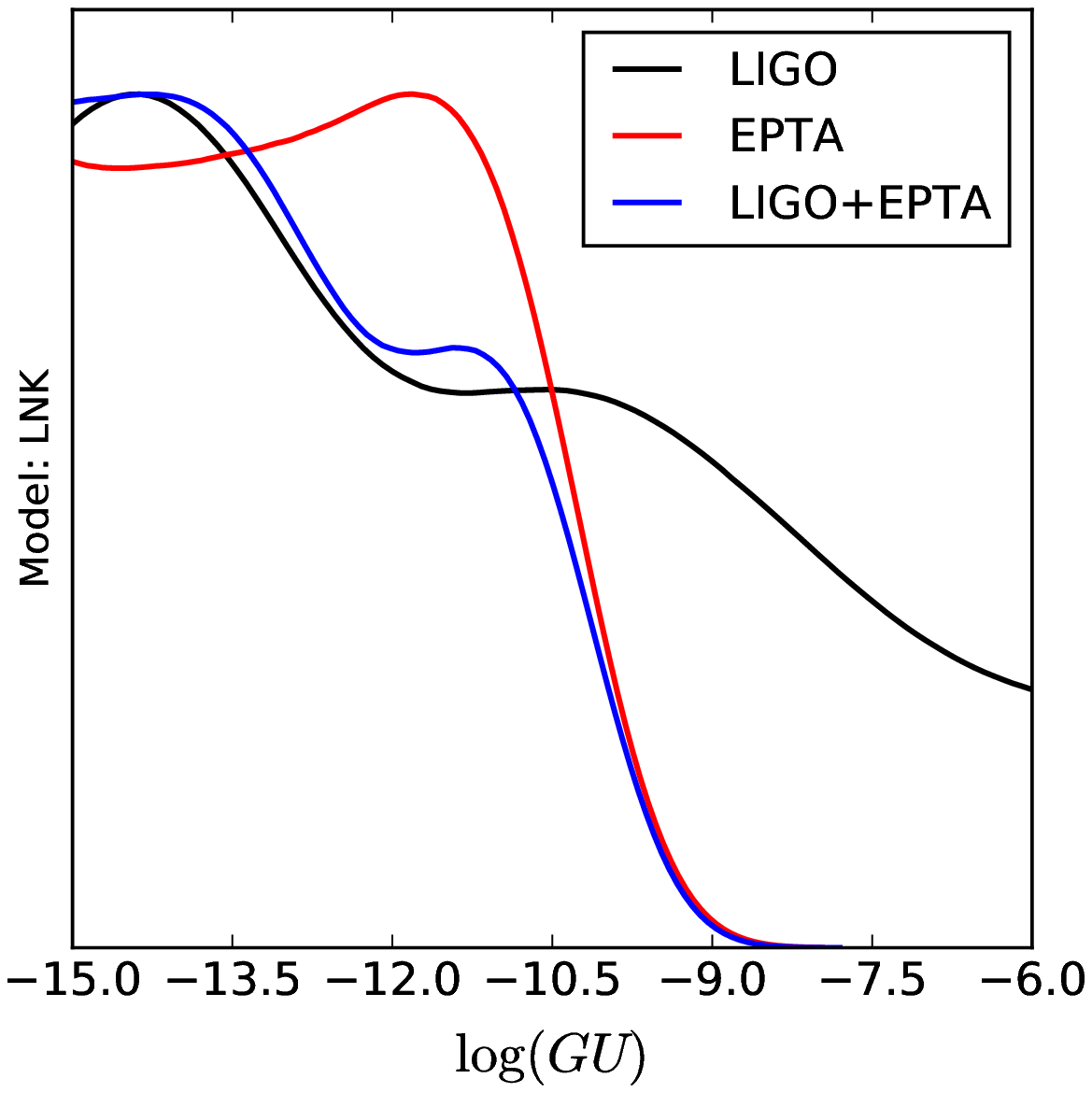}
    \includegraphics[width=\twofigw]{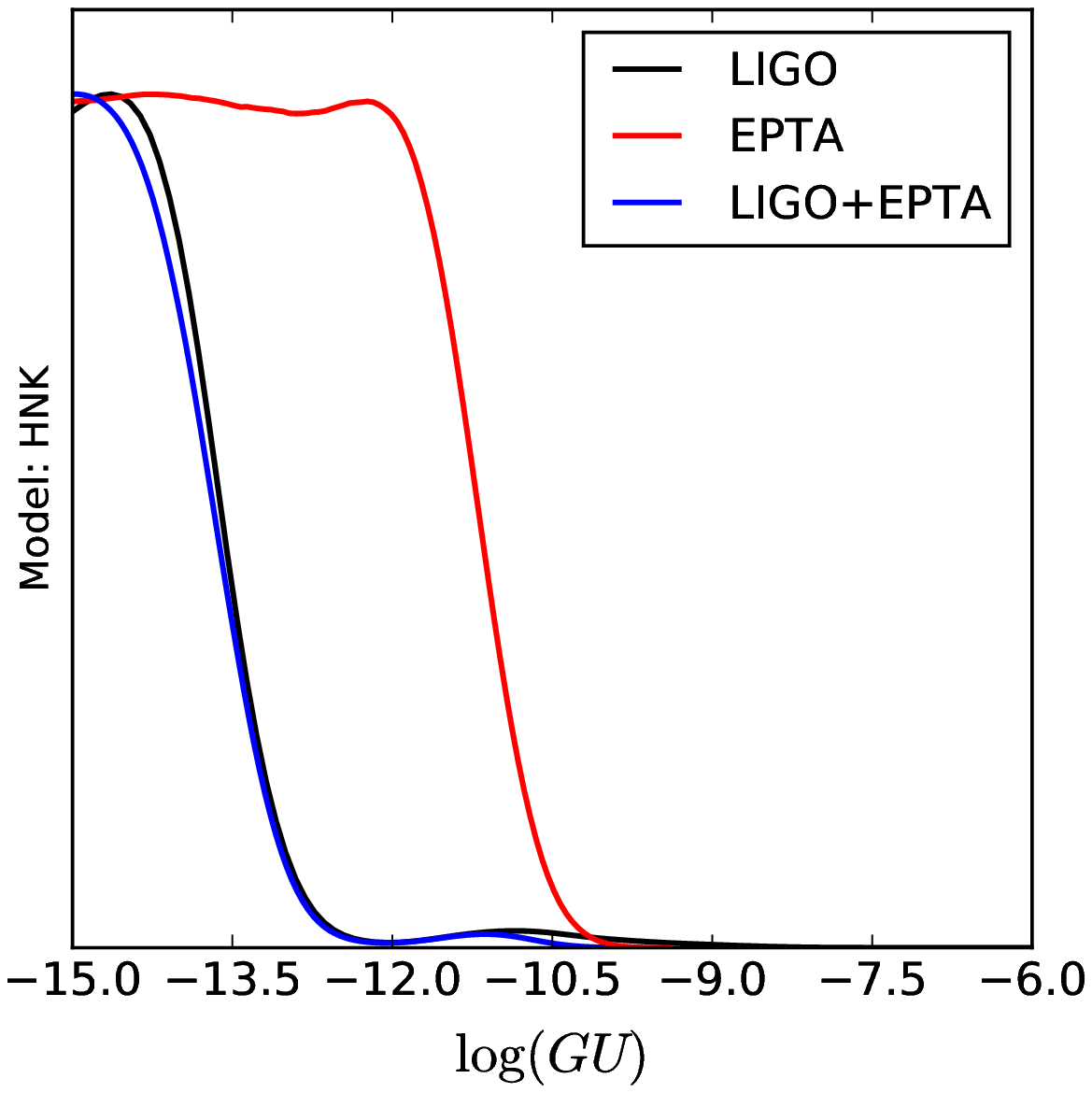}
    \includegraphics[width=\twofigw]{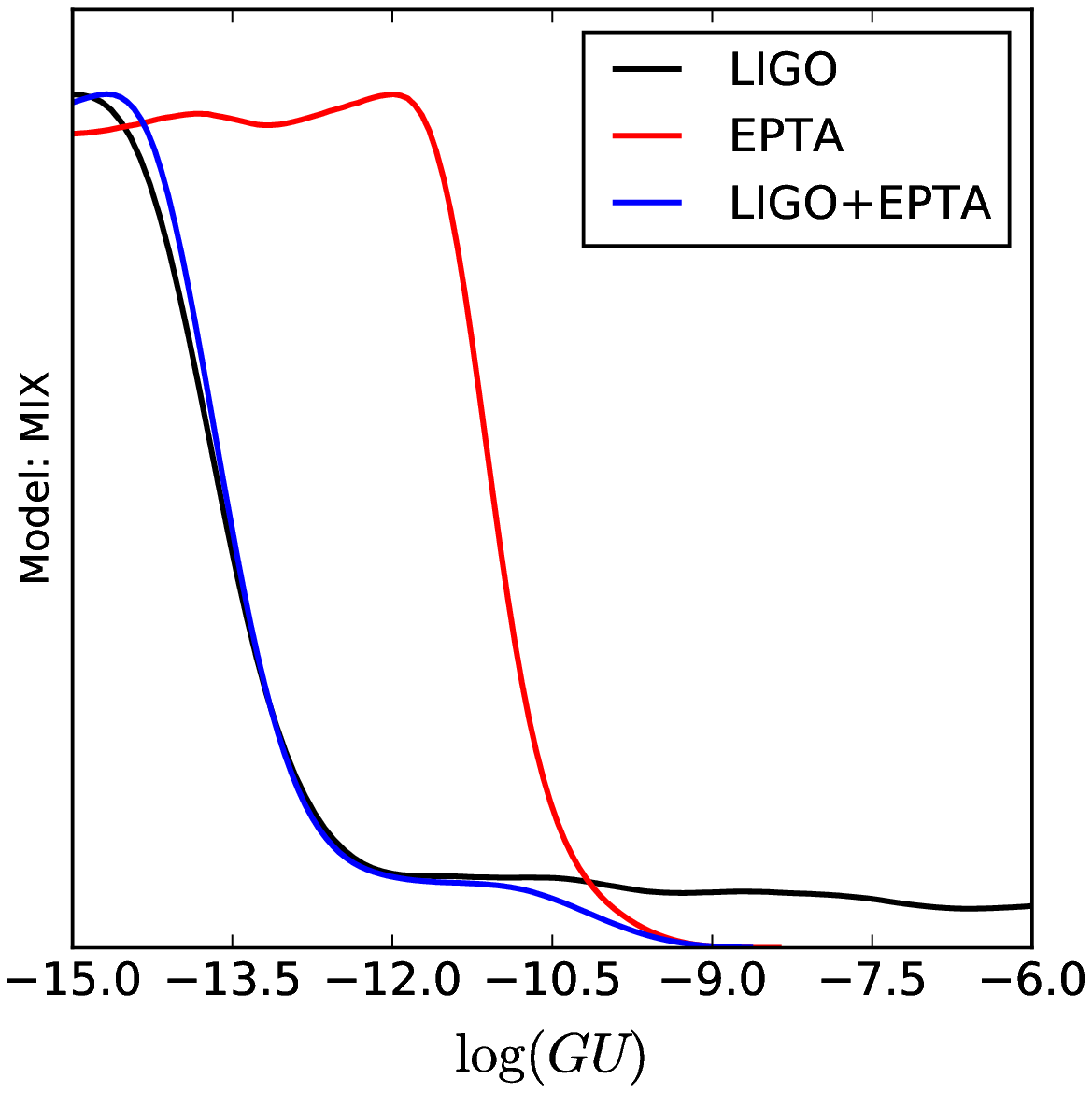}
    \caption{One dimensional marginalized posterior probability
      distribution for the string tension $\GU$ associated with the
      microstructure models 2C, LNK, HNK and MIX. Apart for LNK, and
      to some extend MIX, all the models are strongly constrained by
      both LIGO and EPTA.}
    \label{fig:1D}
  \end{center}
\end{figure}

\subsubsection{LIGO and EPTA data}

Unfortunately, up to our knowledge, none of these works have provided
a public likelihood based on the frequencies that have been sampled to
search for stochastic GW. As a result, we have resorted to a basic
two-dimensional Gaussian likelihood fitting of their two-sigma upper
limit in the plane $(\Omega_\alpha,\alpha)$ where
\begin{equation}
\OmegaSGW(f) \equiv \Omega_\alpha \left(\dfrac{f}{\fref} \right)^\alpha,
\end{equation}
$\fref$ being the reference frequency of the detector. In order to
test our toy likelihood, we have sampled the parameter space
$(\Omega_\alpha,\alpha)$ using the nested sampling algorithm
{\MULTINEST}~\cite{Feroz:2007kg, Trotta:2008bp, Feroz:2008xx,
  Feroz:2013hea} and extracted the one- and two-sigma contours of the
associated two-dimensional posterior distributions. They have been
plotted in figure~\ref{fig:ligoepta_2D} for both the LIGO and EPTA
data. Our likelihood reproduces well the confidence interval plotted
in figure~2 of Ref.~\cite{TheLIGOScientific:2016dpb} and in figure~14
of Ref.~\cite{Verbiest:2016vem}.

Let us stress the importance of keeping the spectral index $\alpha$ in
the inference problem. As can be seen in
figure~\ref{fig:ligoepta_2D}, the two-sigma upper limit on the
amplitude $\Omega_\alpha$ may change by more than two orders of
magnitude according to $\alpha$. This is particularly relevant for
cosmic strings. As one can check in figure~\ref{fig:proto}, the slope
of the spectrum at the LIGO and EPTA frequencies, $\fref=25\,\Hz$ and
$\fref = 31.7\,\nHz$, respectively, depends on all the string
parameters, and in particular $\GU$.

\subsubsection{Constraints on the string tension}

\begin{figure}
  \begin{center}
    \includegraphics[height=\threefigw]{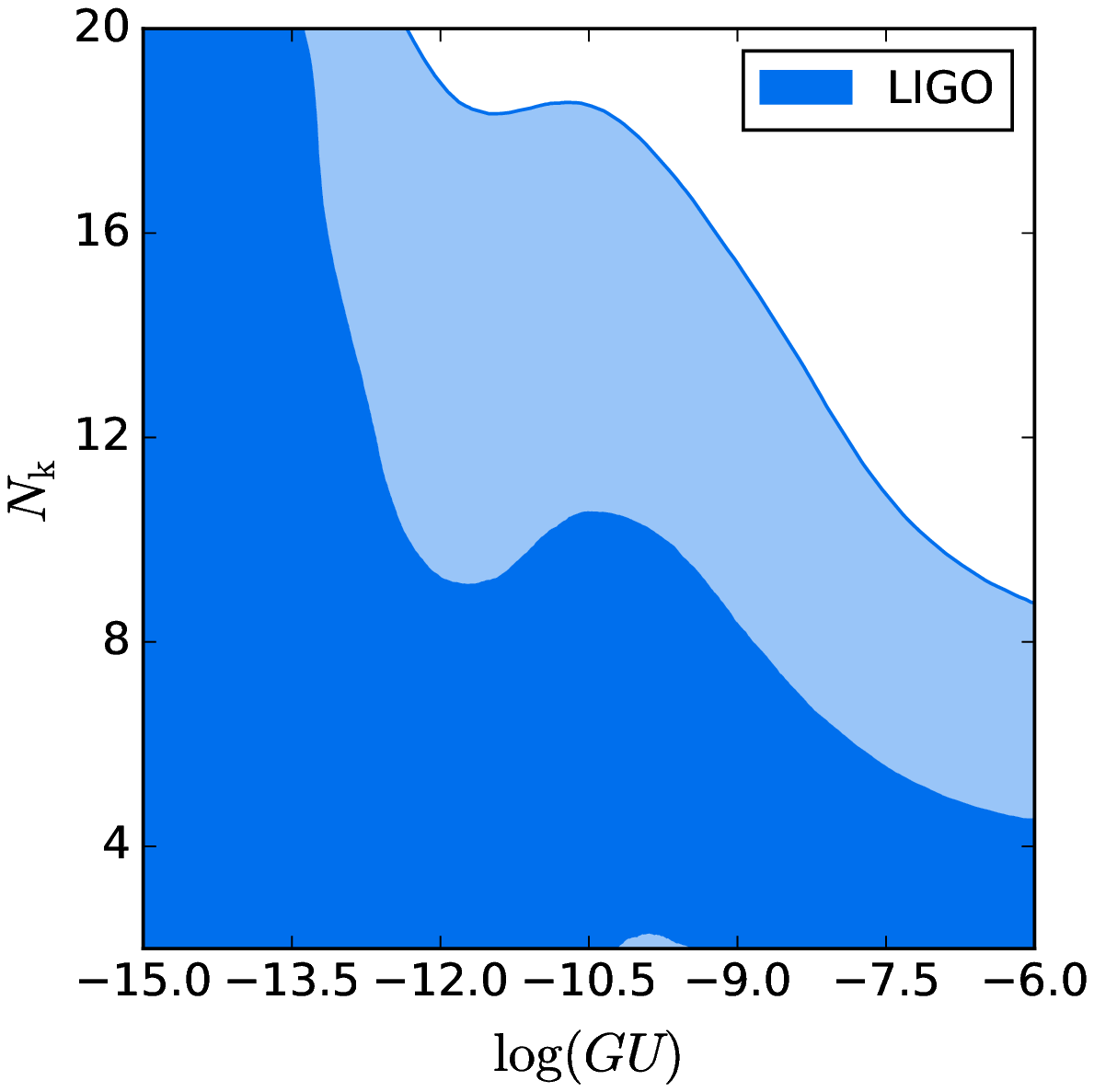}
    \includegraphics[height=\threefigw]{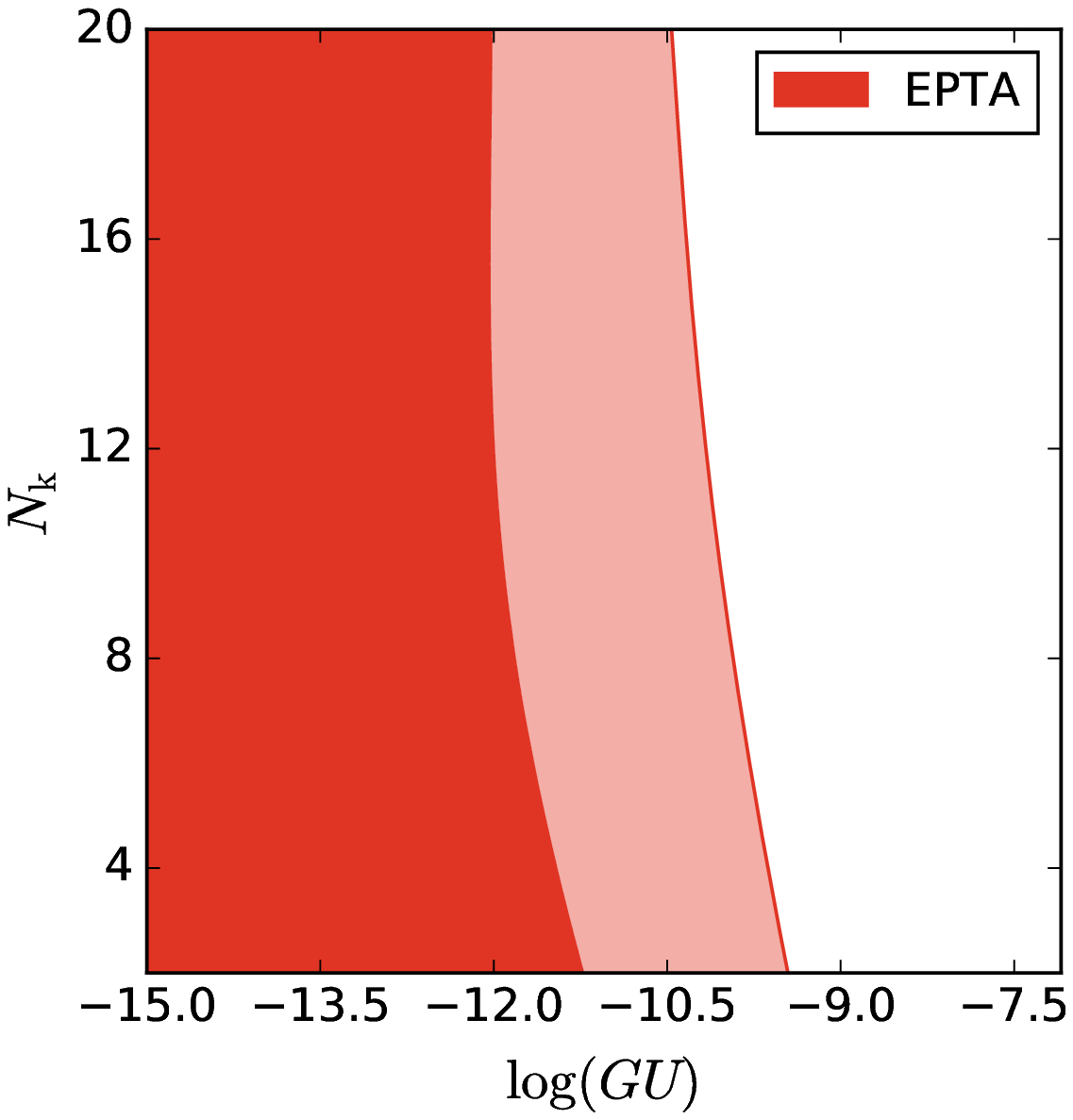}
    \includegraphics[height=\threefigw]{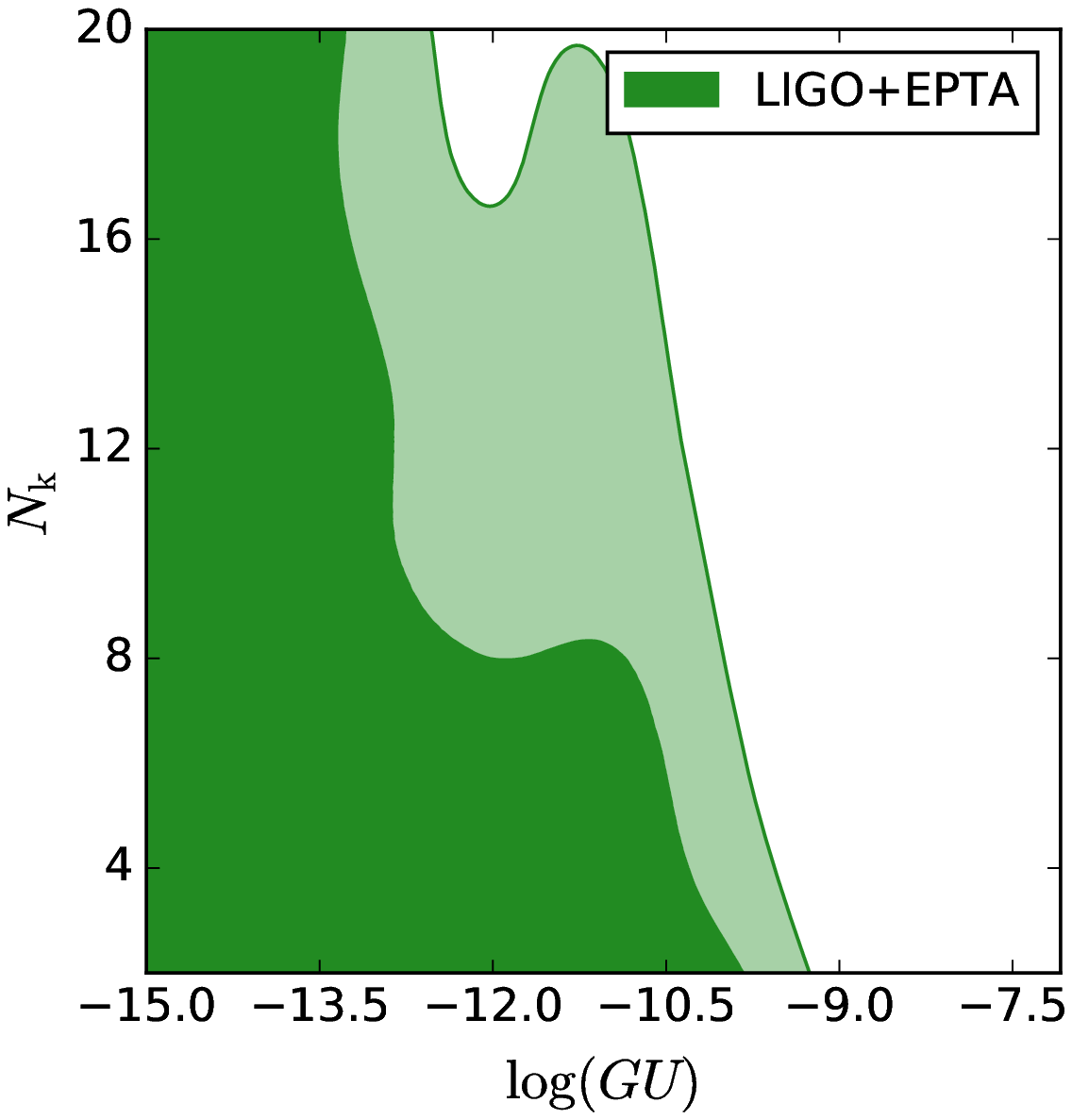}
    \caption{One- and two-sigma contours of the posterior probability
      distributions in the plane $[\Nkink,\log(\GU)]$ for the low
      number of kinks microstructure model LNK. For $\Nkink<8$, there
      is no constraints coming from the LIGO bounds.}
    \label{fig:LNK_2D}
  \end{center}
\end{figure}

\begin{figure}
  \begin{center}
    \includegraphics[height=\threefigw]{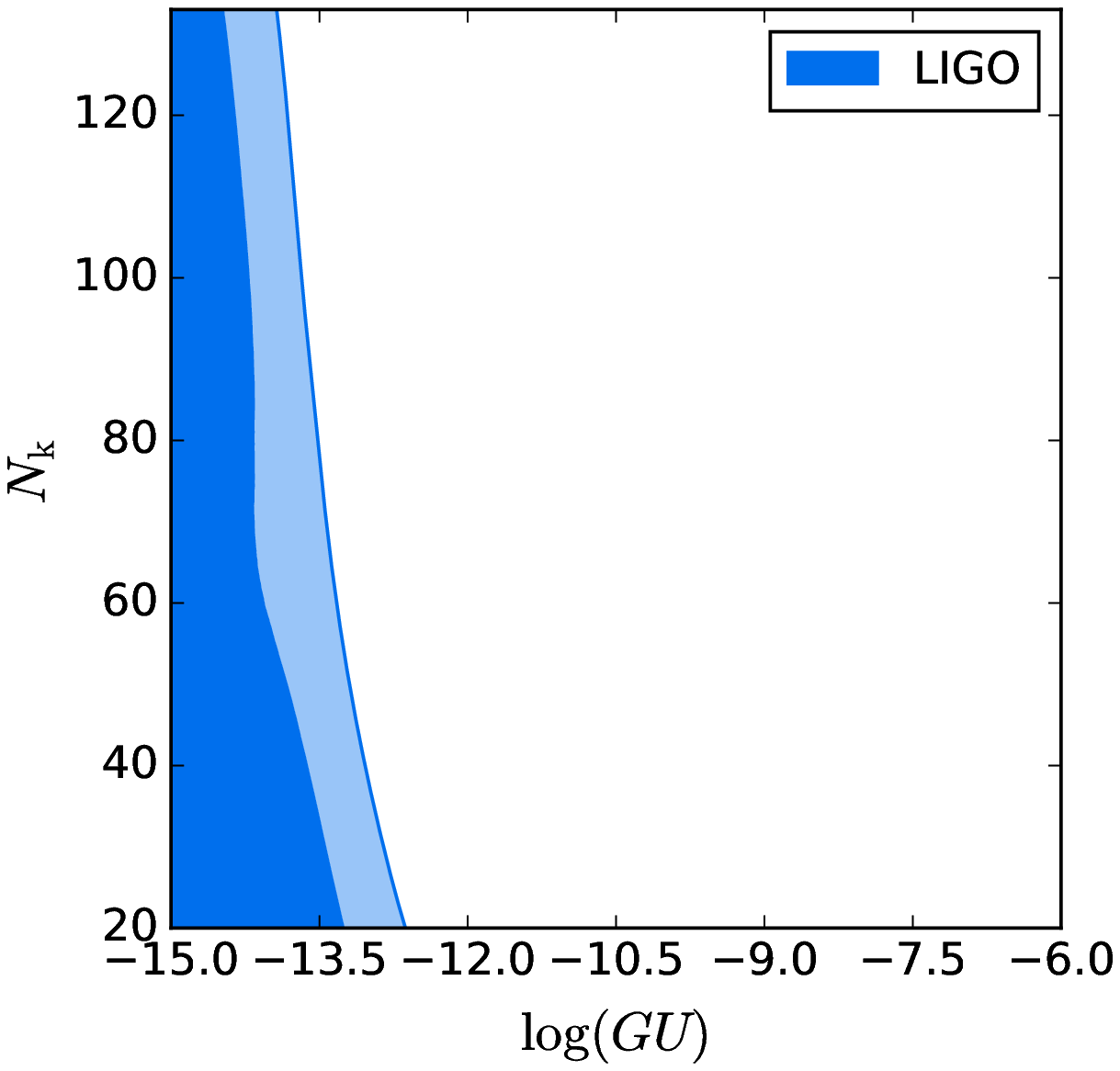}
    \includegraphics[height=\threefigw]{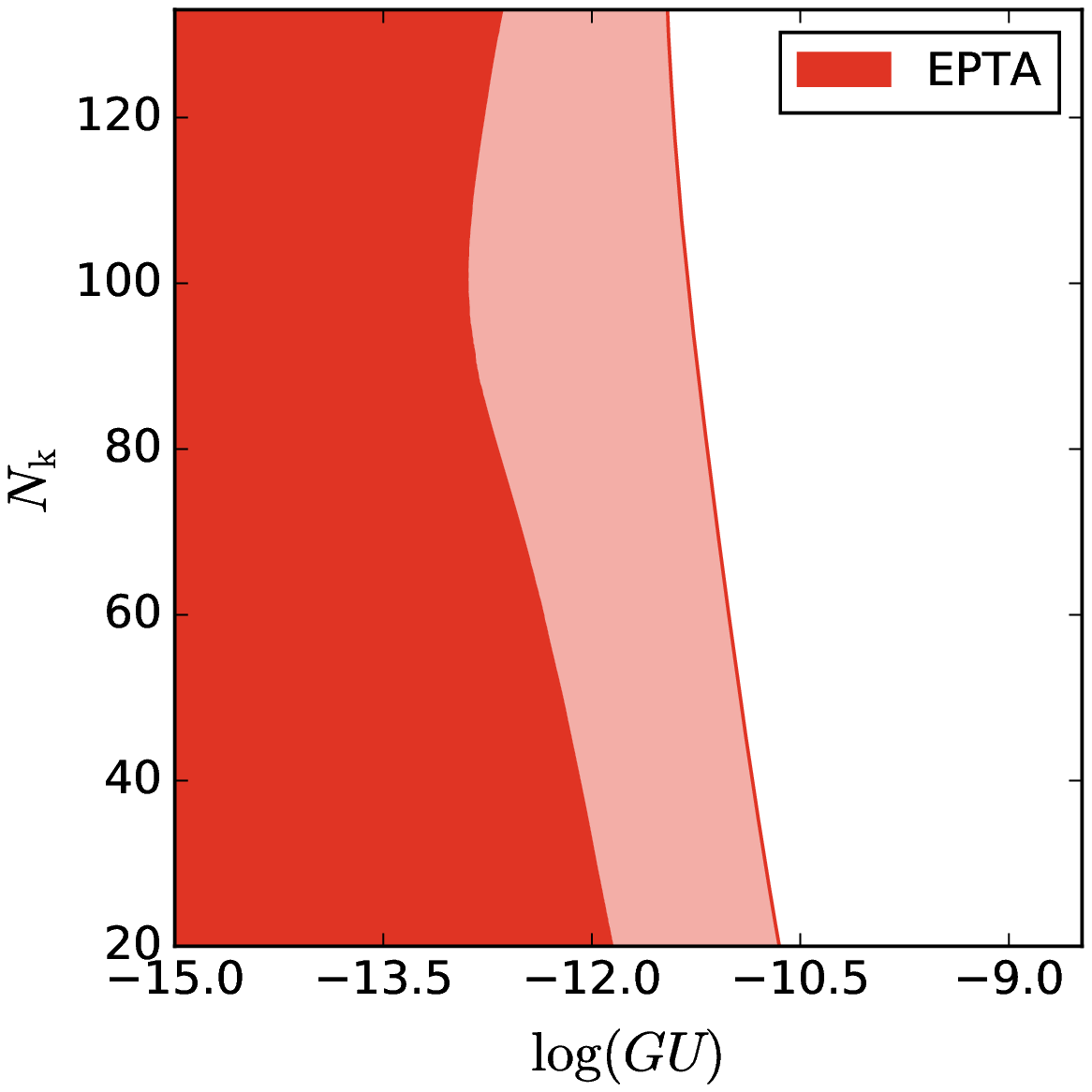}
    \includegraphics[height=\threefigw]{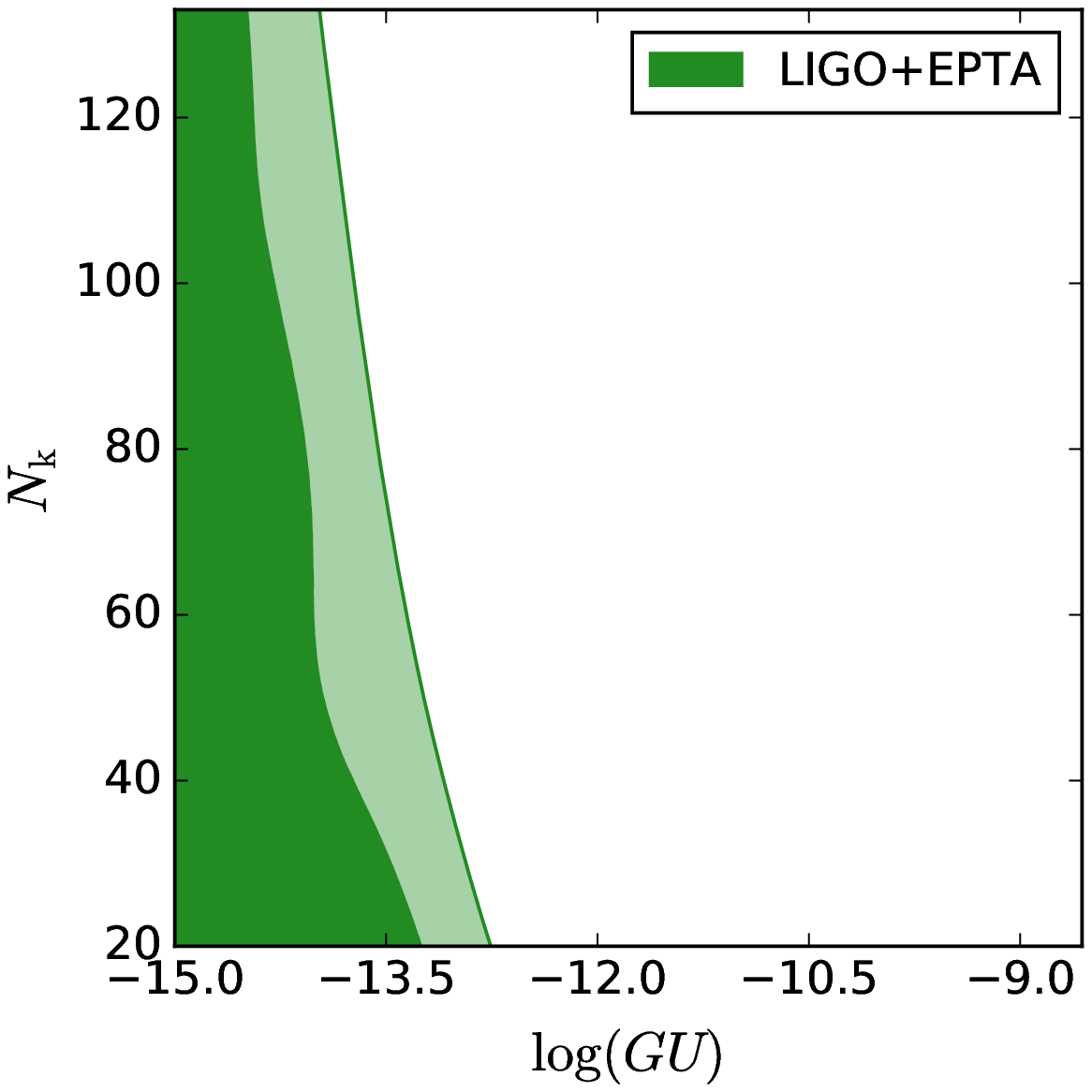}
    \caption{One- and two-sigma contours of the posterior probability distributions in
      the plane $[\Nkink,\log(\GU)]$ for the high number of kinks microstructure
      model HNK. This scenario is severely constrained by LIGO.}
    \label{fig:HNK_2D}
  \end{center}
\end{figure}

\begin{figure}
  \begin{center}
    \includegraphics[height=\threefigw]{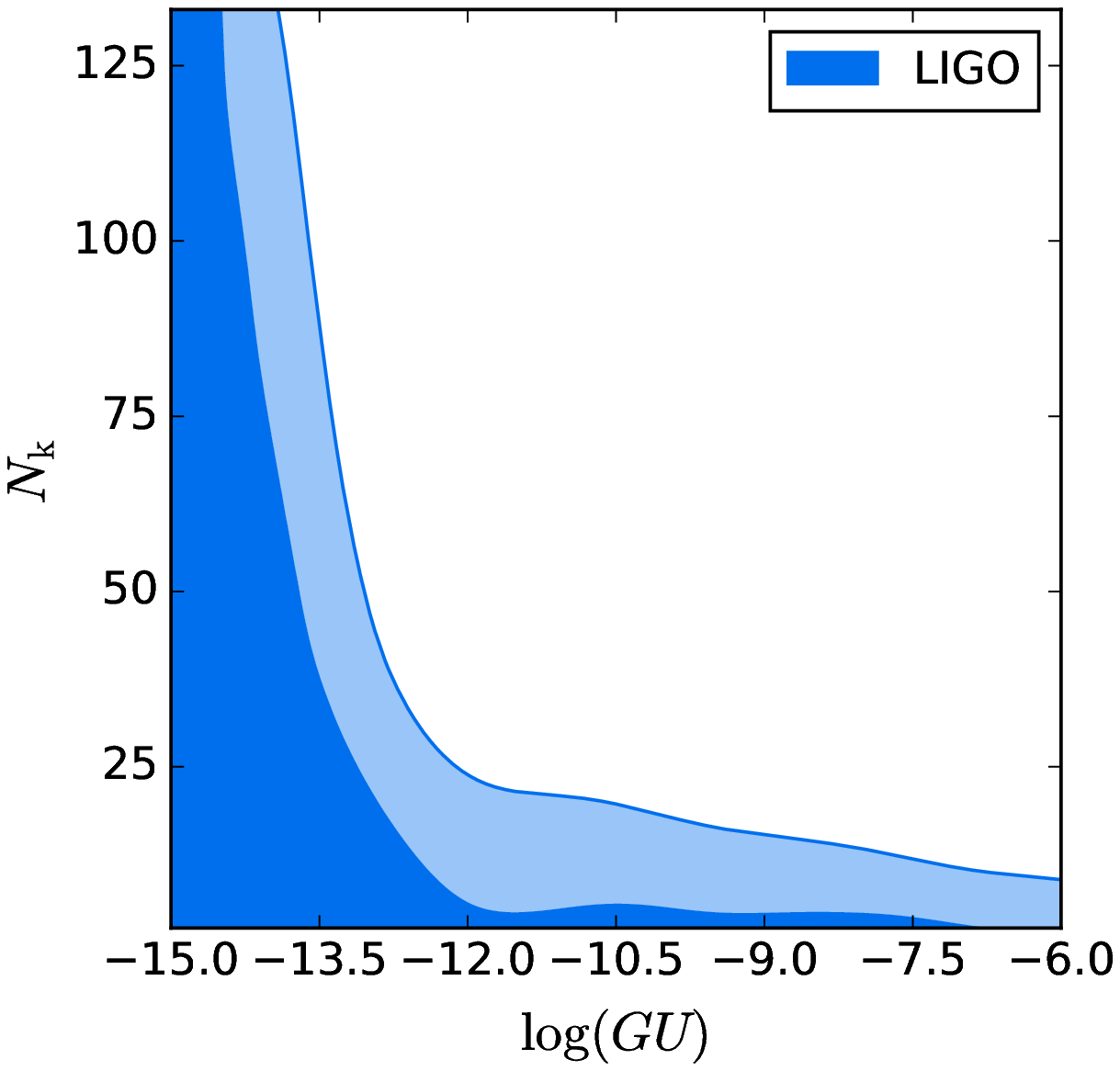}
    \includegraphics[height=\threefigw]{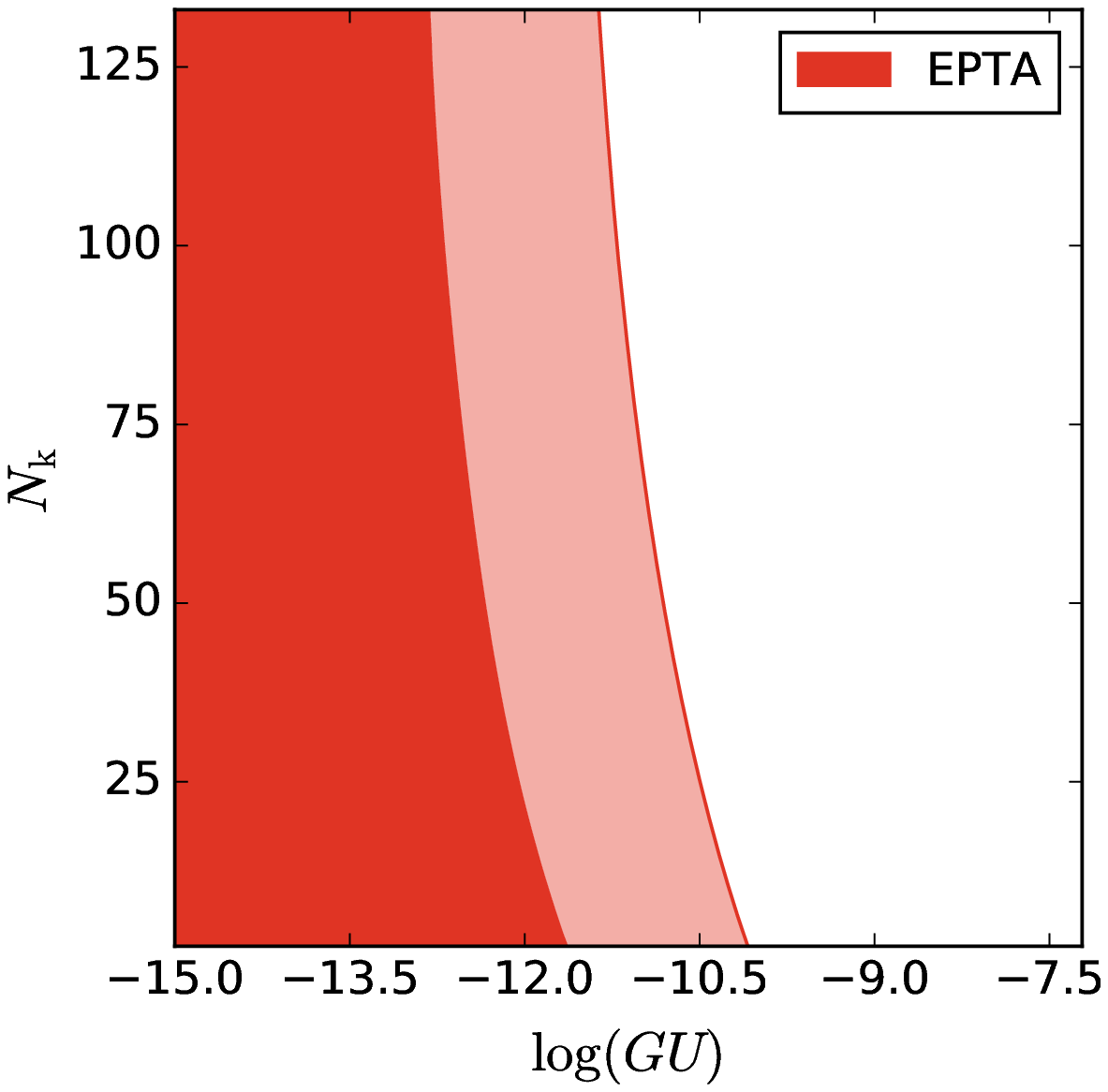}
    \includegraphics[height=\threefigw]{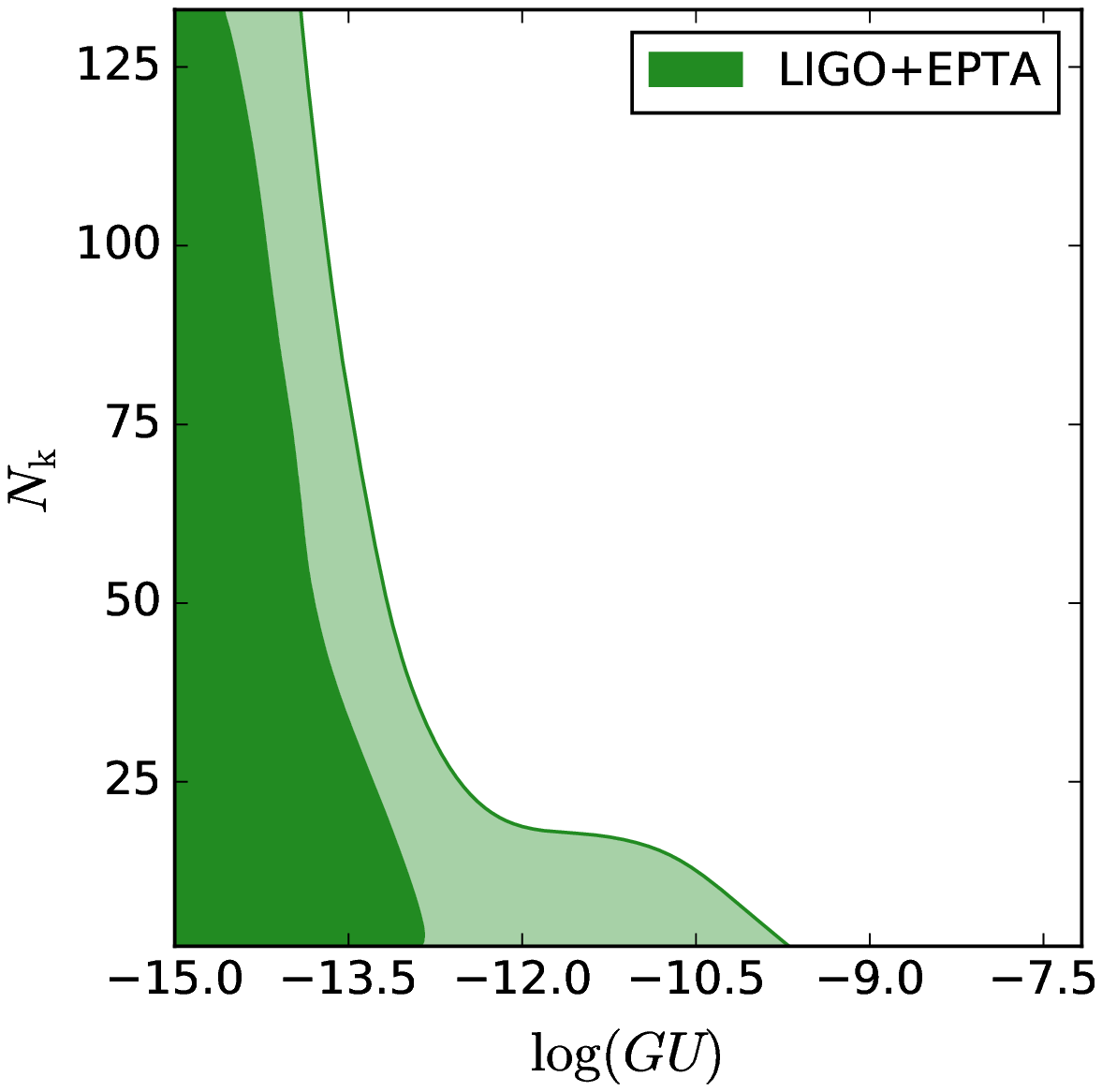}
    \caption{One- and two-sigma contours of the posterior probability
      distributions in the plane $[\Nkink,\log(\GU)]$ for the model
      MIX having only two cusps for very small loops ($\gamma < \gammac$) and
      $\Nkink$ kinks for the larger loops. This scenario is severely
      constrained by EPTA while LIGO is sensitive to values of
      $\Nkink \gg 20$ only.}
    \label{fig:MIX_2D}
  \end{center}
\end{figure}

For the four prototypical models discussed earlier, namely 2C, LNK,
HNK and MIX, we can evaluate both $\Omega_\alpha$ and $\alpha$ at the
two frequencies $\fref=25\,\Hz$ and $\fref = 31.7\,\nHz$,
corresponding to LIGO and EPTA, respectively. Plugging these numbers
into our toy likelihood allows us, in return, to test the ability of
each model to satisfy the LIGO and EPTA bounds.

In practice, we have sampled the parameter space associated with each
model by using the nested sampling algorithm {\MULTINEST}. For
instance, for the model 2C, the only parameter is $\GU$, for which we
have chosen a Jeffreys' prior as $\log(\GU)\in[-15,-6]$. For the other
models, LNK, HNK and MIX, we have kept the same prior for $\GU$ and
considered a flat prior for the number of kinks $\Nkink$ in the ranges
discussed in section~\ref{sec:proto}. Although the dimensionality of
the problem is small, two at maximum, calculating all the cusps, kinks
and collisions part of the spectra amounts for about ten seconds of
computation rendering the whole analysis numerically heavy. These
difficulties have been overcome by using the MPI-parallelized version
of {\MULTINEST} over a $5000$ live points and we have stopped the
sampling for an expected error on the Bayesian evidence at
$10^{-4}$. This typically corresponds to a total number of samples
around $10^4$, which is accurate enough to determine the marginalized
posterior in a two-dimensional parameter
space~\cite{Ringeval:2013lea}.

The one-dimensional posteriors for $\GU$ have been represented in
figure~\ref{fig:1D}. Notice that, for the models LNK, HNK and MIX,
these posteriors are obtained by marginalization over $\Nkink$. The
corresponding one- and two-sigma contours of the two-dimensional
posteriors in the plane $[\log(\GU),\Nkink]$ have been plotted in
figures~\ref{fig:LNK_2D}, \ref{fig:HNK_2D} and \ref{fig:MIX_2D}. Among
other effects discussed below, the bounds on $\GU$ become stronger for
higher values of $\Nkink$, as one may expect.

Let us first notice that we find almost all the models to be in the
sensitivity regime of both LIGO and EPTA. There is indeed a
significant amount of power around the LIGO frequency $\fref =
25\,\Hz$ coming from the peak associated with the gravitational
backreaction length scale, an effect neglected so far. Interestingly
enough, we find that the constraints set by LIGO and EPTA on $\GU$ for
the smooth model 2C are within an order of magnitude range, LIGO being
slightly less constraining than EPTA. The moderately kinky model, LNK,
falls short under the LIGO sensitivity threshold but is well
constrained by EPTA. On the contrary, the very kinky loop model HNK is
strongly constrained by LIGO, and to a lesser extend by EPTA. These
results should make clear the complementarity of using two very
different frequencies. A closer look to figure~\ref{fig:1D} for LIGO
reveals a change of behaviour in the tail of the posterior
distribution. This feature can be understood from
figure~\ref{fig:GUdep}. By lowering $\GU$, the peak of the spectrum
moves to higher frequencies and, combined with thermal history
effects, around $f=10\, \Hz$, this can actually induce a transient
increase in power.

Finally, the MIX model is slightly constrained by LIGO and much
constrained by EPTA. As can be seen in figure~\ref{fig:proto}, the
low-frequency part of the spectrum associated with MIX is dominated by
kinks only. As a result, the EPTA constraint on $\GU$ for MIX ends up
being in between the one associated with LNK and HNK. Around the LIGO
frequencies, cusps are present but their overall contribution to the
spectrum remains smaller than the one coming from kinks and
collisions, at least for most of the possible values of $\Nkink$. In
figure~\ref{fig:1D}, one can see that the tail of the one-dimensional
posterior distribution for $\GU$ is not vanishing. From the
two-dimensional posterior of figure~\ref{fig:MIX_2D}, it comes from
the low values of $\Nkink$ which fall under the sensitivity threshold
of LIGO, as it is the case for LNK. As a result, we find that there is
only a two-sigma LIGO bound on $\GU \le 1.4 \times 10^{-8}$, and no
constraint at all at three-sigma.

\begin{table}
\begin{center}
  \begin{tabular}{|c|l|l|l|}
  \hline
  Model & LIGO & EPTA & LIGO + EPTA \\
  \hline
   2C & $\GU \le 1.1 \times 10^{-10} $ & $\GU \le 3.4 \times 10^{-11}$
   & $\GU \le 1.0 \times 10^{-11}$ \\
   \hline
   LNK & $-$ &$\GU \le 6.8\times 10^{-11}$ & $\GU \le 7.2\times
   10^{-11}$\\
   \hline
   HNK & $\GU \le 8.8 \times 10^{-14}$ & $\GU \le 6.4 \times 10^{-12}$ & $\GU \le 6.7 \times 10^{-14}$ \\
   \hline
   MIX & $\GU \le 1.4 \times 10^{-8}$ & $\GU \le 1.1 \times 10^{-11}$
   & $\GU \le 5.9 \times 10^{-12}$ \\
   \hline
\end{tabular}
\caption{Two-sigma upper limit for the string tension $\GU$ associated
  with the prototypical models of string microstructure. For the two
  cusps model (2C), $\GU$ is the only parameter whereas for the kinky
  models, LNK and HNK, and the scale-dependent model MIX, the
  constraints are obtained by marginalisation over the number of kinks
  $\Nkink$. The EPTA limits apply to all the scenarios whereas LIGO
  ends up being constraining 2C and HNK. For the MIX model, the LIGO
  bound on $\GU$ exists only at two-sigma and disappears at three
  sigma (see text).}
\label{tab:GUlimits}
\end{center}
\end{table}

From the one-dimensional marginalised probability distributions, we
have extracted the two-sigma upper limits on the string tension $\GU$
for each scenario. They have been reported in Table.~\ref{tab:GUlimits}.

\section{Conclusion}

We have provided a new estimation of the stochastic GW spectrum
generated by a network of cosmic strings in scaling. The loop
distribution underlying our calculations comes from Nambu-Goto
simulations and our Boltzmann approach allows us to consider any
relaxation effects that are associated with the transition from the
radiation era to the matter era. Most importantly, we have also
included both the effect of GW emission at $\gamma \le \gammad$ and GW
backreaction at $\gamma \le \gammac$ on the loop distribution. The
fact that these scales do not match yields two characteristic features
in the spectrum which may not have been appreciated enough before.

Because some theoretical uncertainties remain on the loop
microstructure which can survive the cosmological history, we have
considered various motivated possibilities in which cosmic string
loops may, or may not, have a high number of kinks instead of the
usual cusps. In particular, we have shown that kinky loops produce a
stochastic gravitational wave spectrum dominated by the collision of
left- and right-moving kinks, an effect neglected so far. Let us also
mention that, as discussed in Refs.~\cite{Steer:2010jk}, one may also
expect some electromagnetic signatures in this regime.

In order to estimate how much the string microstructure affects the
stochastic GW spectrum, we have devised four prototypical models. The
first (2C) corresponds to loops developing two cusps per oscillation,
it is a smooth string model corresponding to the standard lore. We
have also considered loops having a low number of kinks $\Nkink\le 20$
(LNK) and loops having a higher number of kinks $20<\Nkink \le 133$
(HNK). In the latter case, the overall stochastic GW spectrum is
expected to be dominated by collision events. Another model, MIX, is a
mixture of the 2C, LNK and HNK and describes small smooth loops with two
cusps if their size $\gamma \le \gammac$ whereas all larger loops
$\gamma > \gammac$ are assumed to have only kinks, with $0 \le \Nkink
\le 133$.

The stochastic GW spectrum for these four prototypical scenarios has
been compared to the current LIGO and EPTA bounds. For this purpose,
we have used a nested sampling exploration of the two-dimensional
parameter space associated with the string tension $\GU$ and the
number of kinks $\Nkink$. By marginalizing over the number of kinks,
we have extracted the two-sigma upper limits on the string tension
$\GU$, for each of the scenario. The results have been summarized in
Table.~\ref{tab:GUlimits}. All the scenarios are constrained by the
joined LIGO and EPTA limits on the stochastic GW background, the
smooth and moderately kinky models (MIX included) having $\GU \le
\order{1} \times 10^{-11}$ at two-sigma level while the very kinky one
satisfies $\GU \le \order{1} \times 10^{-14}$. Let us stress again the
complementarity of using both LIGO and EPTA which are probing
different frequency domains. The two cusps (2C), the low number of
kinks (LNK) and the mixture (MIX) models are more constrained by the
EPTA bounds in the low frequency part of the spectrum whereas the high
number of kinks scenario (HNK) is already severely bounded by LIGO
alone. The LIGO frequency domain is indeed matching the peak of the
string-generated GW background when GW backreaction on the loop
distribution is considered. According to
Ref.~\cite{TheLIGOScientific:2016dpb}, the design sensitivity of LIGO
in this frequency could yield a two orders of magnitude improvement on
$\Omega_\alpha$, and thus on $\GU$. Let us also stress that if
stochastic GW end up being detected by LIGO, the frequency range could
be exactly right to the peak, precisely in the region for which the
spectral index of the spectrum is given by the loop
microstructure. One would therefore know if cosmic string loops are
made of cusps, kinks or collisions.

\acknowledgments It is a pleasure to thank Mairi Sakellariadou and
Daniele Steer for providing original motivations for this work as well
as the participants and organisers of the 2015 eLISA Cosmology Working
Group workshop for enlightening discussions.  This work was supported
in part by JSPS Grant-in-Aid for Scientific Research No.~15K17632
(T.~S.), and MEXT KAKENHI Grant Numbers 17H06357 (T.~S.) and 17H06358
(T.~S.).

\section*{Note added}

On the day before this paper was submitted, two other papers appeared
online, \href{https://arxiv.org/abs/1709.02434}{1709.02434} and
\href{https://arxiv.org/abs/1709.02693}{1709.02693}, providing
new constraints on the cosmic string tension $\GU$. They consider only
a smooth loop model, and this would fit within our ``2C'' scenario. Up
to an order one factor, their constraint indeed matches ours for the
Pulsar Timing Array data and the 2C model. This is expected as the low
frequency part of our spectrum is the same as the one of
Ref.~\cite{Blanco-Pillado:2013qja}. However, they do not find any
constraint from LIGO, probably due to their simplifying assumption of
matching the scale of gravitational wave emission and gravitational
wave backreaction.

\bibliographystyle{JHEP}

\bibliography{strings}

\end{document}